\documentclass[pre,amsmath,amssymb,superscriptaddress,letterpaper]{revtex4}

\usepackage{graphicx}% Include figure files
\usepackage{subfigure}

\begin{document}

\title{Channel noise induced stochastic facilitation in an auditory brainstem neuron model}
\author{Brett A. Schmerl}
 \email{brett.schmerl@mymail.unisa.edu.au}
\affiliation{Computational and Theoretical Neuroscience Laboratory, Institute for Telecommunications Research, University of South Australia, SA 5095, Australia}
\author{Mark D. McDonnell}
 \email{mark.mcdonnell@unisa.edu.au}
\affiliation{Computational and Theoretical Neuroscience Laboratory, Institute for Telecommunications Research, University of South Australia, SA 5095, Australia}

\date{\today}

\begin{abstract}

Neuronal membrane potentials fluctuate stochastically due to conductance changes caused by random transitions between the open and close states of ion channels. Although it has previously been shown that channel noise can nontrivially affect neuronal dynamics, it is unknown whether ion-channel noise is strong enough to act as a noise source for  hypothesised noise-enhanced information processing in real neuronal systems, i.e.~`stochastic facilitation.' Here, we demonstrate that biophysical models of channel noise can give rise to two kinds of recently discovered stochastic facilitation effects in a Hodgkin-Huxley-like model of auditory brainstem neurons. The first, known as slope-based stochastic resonance (SBSR), enables phasic neurons to emit action potentials that can encode the slope of inputs that vary slowly relative to key time-constants in the model. The second, known as inverse stochastic resonance (ISR), occurs in tonically firing neurons when small levels of noise  inhibit tonic firing and replace it with burst-like dynamics. Consistent with previous work, we conclude that channel noise can provide significant variability in firing dynamics, even for large numbers of channels. Moreover, our results show that  possible associated computational benefits may occur due to channel noise in neurons of the auditory brainstem. This holds whether the firing dynamics in the model are phasic (SBSR can occur due to channel noise) or tonic (ISR can occur due to channel noise).
\end{abstract}

\maketitle

\section{Introduction}

Despite the ubiquity of numerous forms of stochastic neuronal noise and variability~\cite{Faisal.08,Ermentrout.08,Rolls,Destexhe}, its influence on neural information processing is not fully understood~\cite{McDonnell}.  It has been shown that the presence of stochastic noise can lead to non trivial beneficial effects in neurobiological systems, such as stochastic resonance~\cite{Gammaitoni.98,McDonnell.PLOS09}. However, beyond stochastic resonance, many other stochastic phenomena have been interpreted as potentially providing some benefit to neuronal information processing, and in order to highlight  this variety, the occurrence of {\em all} such effects has been labelled as `stochastic facilitation'~\cite{McDonnell}. For example, classical stochastic resonance is one particular case of this general concept. A key aspect of the `stochastic facilitation' perspective is to link a computational hypothesis with neural dynamics. Stochastic facilitation can be said to occur once a particular neural computation is assumed to be required, and a stochastic dynamical mechanism observed as a plausible means of instantiating it.  There are no restrictions on what that computation might be, provided there is some evidence that a neurobiological system could plausibly perform it. 

Of particular interest to date has been  the effect of noise at the level of single neurons, as effects at this scale are likely to significantly influence higher levels of description, such as the behavior of neuronal populations in networks~\cite{Rolls,Buesing.11}. Neuronal noise resulting from stochastic synaptic vesicle release is perhaps the most frequently studied source of stochastic variability in neuronal systems (see, e.g.~\cite{Tsodyks.97,Lindner.09}). However, another stochastic process that occurs in vivo, {\em channel noise}~\cite{Hille,FoxLu,FOX,Chow,White}, has been shown to influence spike reliability, population responses, and information processing~\cite{Schneidman,Shuai.05,Faisal.07,Rowat.07,Ashida,Goldwynwhat}.  We also note that stochastic resonance due to ion channel noise has been previously studied~\cite{Bezrukov.95,Goychuk.00}. However, our focus here is to try to identify mechanisms that could be useful in spike-based neuronal computation and which rely on intrinsic stochastic phenomena, such as channel noise. To this end, our study is different in three ways to prior work on stochastic resonance in ion channels~\cite{Bezrukov.95,Goychuk.00}. First,  we study the influence of noise at the scale of ion channels on the dynamics of spiking in detailed neuron models, rather than on information processing at the channel scale. Second, we consider new forms of stochastic facilitation that have only recently been described. Third, we demonstrate the utility of efficient channel noise approximation methods for studies of this nature.  

Recent renewed interest in channel noise has led to  derivations of a variety of new computationally efficient methods for simulating it, mainly using diffusion approximations~\cite{Bruce.07,Bruce.09,Sengupta.10,Goldwyn,Goldwynwhat,Linaro,Schmandt,Dangerfield,Orio.12,Huang.13}. The question of how to improve the speed of simulations by employing useful  approximations was also considered many years previously, e.g.~\cite{FOX,FoxLu,Mino02}. Here, we make use of one such method, but our focus is predominantly on the consequences of ion channel noise rather than on any particular model or simulation method.

Channel noise has also been considered an important factor for consideration in how auditory neurons encode information. This has led to the suggestion of utilizing channel noise to reproduce more natural responses in populations of neurons stimulated by cochlear implants~\cite{White,Mino04,Imennov,Goldwyn.12}. Channel noise is particularly important in this example as it is likely to be the most significant biophysical source of stochastic noise in the peripheral auditory system following hearing loss caused by the loss of inner hair cells~\cite{White}.

Channel noise can also be thought of as  a source of intrinsic noise in electrically active cells, as opposed to extrinsic noise, which may arrive from stochastically varying spike train  input, or stochastic variability due to probabilistic synaptic neurotransmitter release~\cite{Linaro}. Although ion channels are complex and membrane fluctuations caused by ion channel dynamics may be a consequence of numerous factors~\cite{Hille}, {\em channel noise} in the sense studied by others, e.g.~\cite{Schneidman,Faisal.07,Ashida,Goldwynwhat}, arises from the stochastic switching between conducting (open) and non-conducting (closed) conformations of ion pores on the cellular membrane. These ion channels determine the conductance of the membrane to a given ion type, and are thus crucial elements in determining a broad repertoire of dynamical neuronal behavior.

In typical models of individual neurons, the large number of  ion channels assumed to be present has led to the assumption that the small fluctuations about the mean conductance caused by the opening and closing of individual ion channels are negligible and may be disregarded. The approximation by a continuous deterministic description is therefore commonly used for computational simplicity and the effects of individual channel fluctuations are lost. However  theoretical and experimental evidence has shown that the presence of this noise cannot always be neglected whilst still achieving an accurate behavioral description, and this has led to this assumption being re-evaluated~\cite{Schneidman, White, Chow,Shuai.05,Rowat.07,Ashida}.  For example, Schneidman \emph{et~al.}~\cite{Schneidman}, highlighted that the magnitude of the fluctuations in the conductances, relative to the mean conductance, is large if the probability of an ion channel being in the conducting state is low. Thus if the activation variables are low and near-threshold, spike timing and reliability may be affected by channel noise, even at large channel numbers~\cite{Schneidman}. Another example analysed by Shuai and Jung is that spontaneous spiking rates exhibit maxima at certain small numbers of ion-channels, with decreases in rates for ion channel numbers between those inducing maxima~\cite{Shuai.05}.

In this paper, we study a versatile model of auditory brainstem neurons introduced by Rothman and Manis~\cite{RothmanManis,RothmanManis2003a,RothmanManis2003b}.  Recently Gai \emph{et~al.}~\cite{Gai09,Gai10} showed that, in the presence of modelled extrinsic noise, a particular case of the Rothman-Manis model exhibited a form of stochastic facilitation, labeled Slope Based Stochastic Resonance (SBSR). The extrinsic noise in this model was introduced to the otherwise deterministic system as white noise added to constant input current, and thus models neurophysiological experiments where noisy current waveforms may be injected into a neuron. It is not clear whether or not the SBSR effect might also be apparent in the presence of biophysically realistic noise such as synaptic or channel noise.

Here, we modify the existing deterministic model of Rothman and Manis by explicitly replacing terms in the models' equations that represent mean conductances with stochastic processes that describe how the actual number of open ion channels fluctuate over time. We simulate the resulting stochastic model using both explicit Monte Carlo simulation of the exact Markov chain model of ion channel states, as well as a system size expansion (SSE) approximation to this system, as recently described by~\cite{Goldwynwhat}. We report that SBSR is indeed observed in the Rothman-Manis model when injected current noise is replaced by modeled channel noise.

Moreover, we show that the channel noise model reproduces a second form of stochastic facilitation, known as Inverse Stochastic Resonance (ISR), in which the presence of a very small amount of noise has the effect of strongly inhibiting the mean spike rate of a model neuron~\cite{Gutkin,Tuckwell.09,Tuckwell.10,Tuckwell}. Unlike  stochastic resonance as it is usually understood~\cite{McDonnell.PLOS09}, in ISR an optimal level of noise does not enhance the system's response to a particular input signal. Instead, the utility of stochastic noise might be hypothesised as important for computational mechanisms that require inhibition of tonic spiking when inhibitory neuromodulation is not available, or alternatively when other computational mechanisms might require on-off bursts of tonic spiking. Such an effect has been observed experimentally~\cite{Paydarfar.06}. Therefore, ISR can clearly be labelled as a form of stochastic facilitation~\cite{McDonnell}, but should not be confused with stochastic resonance~\cite{McDonnell.PLOS09}. 

ISR was originally reported in a Hodgkin-Huxley neuron model and shown to be present with either a synaptic noise model or with white noise added to a deterministic input current~\cite{Gutkin}. Here, we report that the  noise amplitudes required to show the strongest inhibitory effects correspond to  numbers of ion channels of the order of $10^5$ in both the Hodgkin-Huxley and Rothman-Manis Type I-II models. We also found that deterministic-like effects at small noise levels require biologically unrealistically large numbers of channels (in the order of $10^7$). This casts doubt on whether sustained tonic firing observed in the corresponding deterministic model would be observed, should constant current be injected into a real neuron for which the deterministic part of its dynamics are otherwise accurately captured by the model. This conclusion could well be attributed to the use of a simple point neuron model rather than one with spatial extent, or a model where other details are included.

The paper is organised as follows. In Section~\ref{s:Model} we state the existing model of Rothman and Manis, and describe how we introduce stochastic channel noise  into the model. Then in Section~\ref{S:Results} we present simulation results that illustrate that  channel noise  enables both SBSR and ISR to be observed in the model. Finally, Section~\ref{S:Discussion} provides some discussion of our results and conclusions that may be drawn from this study.

A Matlab implementation of the SSE method of simulating
channel noise in the Rothman-Manis model is available online
in the ModelDB repository (accession number 151483, at http://senselab.med.yale.edu/ModelDB/ShowModel.asp?model=151483).
The code can be adapted to any neuron model that can be
expressed in the form of Eq. (1), and specifically implements
Eqs. (8), (9), and (11).

\section{Neuron and Channel Noise Models}\label{s:Model}

\subsection{Generic Hodgkin-Huxley type model}

The form of all models we consider is a generalisation of the Hodgkin-Huxley model (see, for example,~\cite{HH}). It is described by a differential equation of the following form
\begin{align}\label{MembraneEqn}
C_{\rm m} \frac{dV(t)}{dt} = I(t)-\sum_{\rm \{a\}}\bar{g}_{\rm a}G_{\rm a}(t)(V(t)-E_{\rm a}),
\end{align}
where ${C_{\rm m}}$ is the membrane capacitance ({pF}), $V(t)$ is the membrane potential (mV), ${\bar{g}_{\rm a}}$ is the maximum total channel conductance for ion channels of type a, ${E_{\rm a}}$ (mV) is the reversal potential for ion channels of type a, and $I(t)$ represents a constant injected current.  Each term in the sum represents a current, $I_{\rm a}(t)=\bar{g}_{\rm a}G_{\rm a}(t)(V(t)-E_{\rm a})$, since the function $G_{\rm a}(t)$ is dimensionless and constrained to the interval $[0,1]$; this function represents the time course of a conductance.  For a passive leak current (a=lk) we have $G_{\rm lk}(t) = 1$ in this formulation.

In deterministic models, each conductance variable, $G_{\rm a}(t)$ is formed from the product of at least one `activation variable' (or `subunit') and each such variable, $x$,  is described by a differential equation representing ion channel kinetics, of the form
\begin{equation}\label{rates}
    \frac{dx(t)}{dt}=\frac{x_{\infty}(V)-x(t)}{\tau_{x}(V)},
\end{equation}
where
  \begin{equation}\label{tau}
    \tau_{x}(V)=\frac{1}{\alpha_{x}(V)+\beta_{x}(V)}
\end{equation}
and
  \begin{equation}\label{inf}
    x_{\infty}(V)=\frac{\alpha_{x}(V)}{\alpha_{x}(V)+\beta_{x}(V)}.
\end{equation}
In these equations, ${\alpha_{x}}(V)$ and ${\beta_{x}}(V)$ are  voltage dependent rates for the opening and closing of ion channel subunits respectively, as in the standard Hodgkin-Huxley formalism.

For example, the standard Hodgkin-Huxley model~\cite{HH} has two ionic currents, Sodium (a=Na) and Potassium (a=K), and three activation variables, $x=m$, $x=n$ and $x=h$, such that $G_{\rm Na}(t)=m^3h$ and $G_{\rm K}(t)=n^4$. The  equations for $\alpha_{\rm Na}$, $\beta_{\rm Na}$, $\alpha_{\rm K}$ and $\beta_{\rm K}$ are well-known~\cite{HH}. 

Later we will replace the standard deterministic form of each $G_{\rm a}(t)$ with stochastic processes that model the fluctuations induced by random opening and closing of a finite number of ion channels.

\subsection{Auditory Neuron Model}

The model of ventral cochlear nucleus (VCN) neurons developed by Rothman and Manis~\cite{RothmanManis,RothmanManis2003a,RothmanManis2003b} is of the form of Eqn.~(\ref{MembraneEqn}). Unlike the standard Hodgkin-Huxley equation, it is instead comprised of three distinct potassium currents, a sodium current and a hyperpolarization-activated cation current, all of which have been found to be present in VCN neurons. These currents were isolated via the voltage clamp technique and kinetic analysis~\cite{RothmanManis}.  This allowed the kinetic schemes of the ion channels responsible for activation and inactivation of the given currents to be experimentally determined for the potassium currents and verified for the others~\cite{RothmanManis}. Such a model scheme provides an ideal framework in which to study biophysically realistic channel noise beyond the standard Hodgkin-Huxley model and with potential applications to studying auditory neurons.

The model presented in~\cite{RothmanManis2003b} is given in five different configurations corresponding to different cell types. Of these, the Type II configuration shows a classic {\em phasic} response upon perturbation, i.e.~it produces action potentials only at onset of a stimulus change~\cite{Izhikevich.04}. This configuration has been used to model bushy cells in the VCN~\cite{Rothman93,ManisMarx}, and favorably compared to experimental recordings of phasic MSO neurons in gerbils~\cite{Gai09}. It was this configuration of the Rothman-Manis neuron model that was shown to exhibit SBSR by Gai {\em et~al.}~\cite{Gai10}. A phasic neuron  is one that produces a single spike at the onset of a rapidly changing input, but typically does not fire at all for inputs that vary slowly relative to the action-potential generating time-constants of the model. Gai {\em et al.}~also noted that the model fires in response to a certain range of {\em slopes} of the input rather than a range of amplitudes~\cite{Gai09,Gai10}, meaning that this neuron has a threshold defined by the magnitude of the derivative of the input current.

The particular Rothman-Manis model we use consists of the following ionic currents: `fast sodium' (${I_{\rm Na}}$), `high threshold potassium' (${I_{\rm KHT}}$),  `low threshold potassium' (${I_{\rm KLT}}$), `hyperpolarization-activated cation current' (${I_{\rm h}}$) and leak current (${I_{\rm lk}}$). Note that in~\cite{RothmanManis2003b}, a third potassium current---`fast transient' K+ $(I_{\rm A})$---is discussed, but this is not required in the model configurations we use here.

The resulting model requires 6 activation variables, $m,h,n,p,w,z$ and $r$. The particular products that form the conductance time courses, and corresponding constants are shown in Table 1, as are the parameter values of the peak conductances, reversal potentials, leak current and capacitance that we use in the Results section below. The Type I-c, Type I-II and Type II configurations of the model as in~\cite{RothmanManis2003b} are used in this study. These have been adjusted to account for temperature as in~\cite{Gai09}. The functions describing the kinetics of each activation variable are given in Appendix A.

\subsection{Channel Noise Models}\label{sec:CNM}

In the deterministic model given by Equation~(\ref{MembraneEqn}), each conductance time course, $G_{\rm a}(t)$, corresponding to ion channel type a, physically represents the mean fraction of open channels of that type. Mathematically, this is equivalent to the probability that a single ion channel is open (and hence in the conducting state).

In turn, the activation variables are representative of the fraction of ion channel {\em subunits} that are in the conducting (open) state. For example, the combined quantities of $m^3h$ and ${w^{4}z}$ (see Table 1) are representative of the Na$^+$ and K$^+$ channels having individually gated subunits for activation and inactivation, with there being three of type $m$ and four of type $w$. Subunit state transition diagrams  can be found, for  example, in~\cite{Rowat.07,Sengupta.10}.

In order to introduce a realistic model of channel noise, fluctuations in the fraction of open channels need to be modeled explicitly. As in~\cite{Goldwynwhat}, we write the fraction of open channels as the sum of a mean value, and a stochastic term, i.e.
\begin{equation}\label{G_a}
G_{\rm a}(t) = G_{\rm a,det}(t) + \xi_{\rm a}(t),
\end{equation}
where $G_{\rm a,det}$ is the deterministic part of the conductance time course, and is equal to the mean fraction of open channels. For the Rothman-Manis model, $G_{\rm a,det}(t)$ is  listed in Table 1 for each channel type.  This mean is equivalent to the probability that a channel is open at time $t$, given activation variables $\alpha(t)$ and $\beta(t)$. The term $\xi_{\rm a}(t)$ represents channel noise; in an ensemble, the channel noise will be conditionally independent in all neurons, given the membrane potential in each neuron.

We now describe two existing stochastic models of channel noise. The first is the standard Markov model that accurately reflects conceptual models of ion channel subunit openings and closings. The second model is an approximation to the first model that enables greatly reduced simulation runtime.

To begin, we introduce $N_{\rm a}$ to represent the total number of channels of  type a. The most conceptually simple way of incorporating channel noise into neuron models consists of tracking the state of each of the $N_{\rm a}$ channels of each type. The probability that a channel will change state is determined by the voltage dependent transition rates $\alpha_{\rm x}(V)$ and $\beta_{\rm x}(V)$~\cite{dayan}---see Eqns~(\ref{rates})--(\ref{inf}). The transition between states in a channel are assumed to follow a Markov process and are therefore memoryless.   In such a scheme, an open channel occurs when all subunits of a particular ion channel type are in the open state. For example, a sodium channel is in the open state when both the $m$ variable is in state $3$, and the $h$ variable is in state $1$.  We introduce the random variable $\mathcal{N}_{\rm a,o}(t)$ to represent the number of open channels of type a at time $t$. We therefore have
\begin{equation}\label{Markov}
G_{\rm a}(t) = \frac{\mathcal{N}_{\rm a,o}(t)}{N_{\rm a}}.
\end{equation}
Thus, from Eqn.~(\ref{G_a}), the fluctuations in this model can be expressed as
\begin{equation}\label{xi_Markov}
\xi_{\rm a}(t) = \frac{\mathcal{N}_{\rm a,o}(t)}{N_{\rm a}} - G_{\rm a,det}(t).
\end{equation}

Direct computational simulation of this Markov chain model is an accurate method, but it is computationally expensive, since it requires numerous random numbers to be generated at each time step, as well as tracking the state of $N_{\rm a}$ channels for each ion channel type.

Recently, there has been considerable interest in developing new methods capable of approximating the full Markov chain description accurately, but which may be implemented in much faster simulations~\cite{Goldwyn,Goldwynwhat,Linaro,Schmandt,Dangerfield,Orio.12,Huang.13}. These methods generally represent channel noise using stochastic differential equations (SDEs) that statistically approximate the fluctuations present in the Markov chain description by adding derived noise terms to the deterministic system of equations. In a recent review~\cite{Goldwynwhat}, it was concluded that a simulation method based on much earlier mathematical theory due to~\cite{FoxLu} is accurate and efficient. Therefore, here we use that method in the form described by~\cite{Goldwynwhat}, and compare its performance when applied to the problem of identifying SBSR, versus simulation of the Markov model. However, for our ISR studies, although we found that for small numbers of ion channels that the Markov model is consistent with the SDE approximation, a very large number of channels is needed for the effect to be observable. Consequently, a complete comparison of the Markov and SDE approaches for ISR was computationally infeasible, and our results are confined to the SDE approach.

A full treatment of the SDE method can be found in~\cite{FoxLu,Goldwyn,Goldwynwhat} and the references therein. Briefly, the method determines the deterministic component of the conductance time course in the same way as the original deterministic model, but also derives approximations to the noise term, $\xi_{\rm a}(t)$, based on a system size expansion (SSE) of the Markov chain (see~\cite{FOX,FoxLu,Pakdaman}). The noise for each channel type is assumed to be a time-varying Gaussian process that is dependent on the associated activation variables, and therefore also on the membrane potential. Following the approach described in~\cite{Goldwynwhat} we obtained $\xi_{\rm a}(t)$ by solving a set of $M_{\rm a}$ coupled SDEs for  ion channel type a with $M_{\rm a}$ states, of the following form
\begin{equation}\label{SDE}
d{\bf x}_{\rm a} = {\bf A}_{\rm a}{\bf x}_{\rm a}dt+{\bf S}_{\rm a}d{\bf W}_{\rm a},
\end{equation}
where ${\bf x}_{\rm a}$ is a length-$M_{\rm a}$ vector, ${\bf A}_{\rm a}$ and ${\bf D}_{\rm a}:={\bf S}_{\rm a}^2$ are $M_{\rm a}\times M_{\rm a}$ voltage-dependent drift and diffusion matrices respectively. Noise enters the equations via ${\bf W}_{\rm a}$, which is a vector of $M_{\rm a}$ independent standard Brownian  processes. The desired outcome $\xi_{\rm a}(t)$ is given by the $M_{\rm a}$--th element of the vector ${\bf x}_{\rm a}$. Once this is found, the membrane potential equation is given by

\begin{align}\label{MembraneEqn2}
C_{\rm m} \frac{dV(t)}{dt} = I(t)-\sum_{\rm \{a\}}\bar{g}_{\rm a}(G_{\rm a,det}(t)+\xi_{\rm a}(t))(V(t)-E_{\rm a}),
\end{align}

Solving~(\ref{SDE}) requires first finding ${\bf D}_{\rm a}$, and then obtaining its matrix square-root. The drift and diffusion matrices may be constructed for an arbitrary kinetic scheme with $M_{\rm a}$ states, comprised from multiple activation variables (e.g. $m$ and $h$ for sodium channels in the Hodgkin-Huxley model) as follows. The  $M_{\rm a}\times M_{\rm a}$  drift matrix ${\bf A}_{\rm a}$ is identical to the transition matrix from the master equation representation of the Markov chain, as noted by~\cite{Goldwyn}, i.e.
\begin{equation}\label{Master}
\frac{d{\bf p}_{\rm a}}{dt}={\bf A}_{\rm a}{\bf p}_{\rm a},
\end{equation}
where the $M_{\rm a}\times 1$ vector ${\bf p}_{\rm a}$ corresponds to the state occupancy probabilities~\cite{dayan}. These probabilities can be determined at time $t$ from the activation variables (e.g. $m$ and $h$ for a potassium channel), and thus the $M_{\rm a}$--th element of ${\bf p}_{\rm a}$ is equal to the probability that a channel of type a is open (conducting).

In~\cite{Goldwynwhat}, the diffusion matrix ${\bf D}_{\rm a}$ is written for the specific cases of sodium and potassium channels in the Hodgkin-Huxley model.   However, here we provide a general method for constructing ${\bf D}_{\rm a}$ from any given drift matrix ${\bf A}_{\rm a}$, and the number of channels of type a, $N_{\rm a}$. The relationship can be expressed as
\begin{equation}\label{Diff}
N_{\rm a}{\bf D}_{\rm a} = ({\bf A}_{\rm a}{\bf x}_{\rm a}{\bf 1}_{M\times 1}^{\top})\circ {\bf I}_{M\times M}-{\bf A}_{\rm a}\circ ({\bf 1}_{M\times 1}{\bf x}_{\rm a}^{\top})-{\bf A}_{\rm a}^{\top}\circ ({\bf x}_{\rm a}{\bf 1}_{M\times 1}^{\top}),
\end{equation}
where ${\bf 1}_{M\times1}$ is a $M_{\rm a}\times 1$ column vector with all entries equal to $1$, ${\bf I}_{M\times M}$ is the $M_{\rm a}\times M_{\rm a}$ identity matrix, and $\circ$ indicates the Hadamard operator (term by term multiplication). Note that all rows and columns of ${\bf D}_{\rm a}$ sum to zero, and ${\bf D}_{\rm a}^{\top}={\bf D}_{\rm a}$.

Finally, although Eqn~(\ref{Diff}) shows the theoretical dependence of  ${\bf D}_{\rm a}$ on ${\bf x}_{\rm a}$, in practical simulations it is useful to replace the stochastic vector ${\bf x}_{\rm a}$ with ${\bf p}_{\rm a}$, to ensure matrix square roots of ${\bf D}_{\rm a}$ can be found, as discussed in~\cite{Goldwynwhat}. 

\section{Demonstration of stochastic facilitation due to ion channel noise}\label{S:Results}

In this section, we present results based on stochastic simulations of the Markov and SSE models described above. For the Markov model, the states of all ion channels were tracked, and state transitions occurred during simulation interval $t+\Delta t$ if the probability of a transition at time $t$ exceeded a uniformly distributed random number between zero and unity.  All such random numbers were independently generated.
In the system size model, the resulting SDEs were solved using the Euler-Maruyama method~\cite{Kloeden} with a time step of $0.01$~ms, and required generation of independent Gaussian random numbers with zero mean and unit variance.

For the results presented here, we set $N_{\rm a}$ to a common value for all channel types, since our primary motivation is to examine whether stochastic facilitation effects observed with current noise could in principle be exhibited with channel noise. We discuss the limitations of this assumption in the concluding section of the paper. The stochastic effects presented here occur for a large range of channel numbers.

\subsection{Slope Based Stochastic Resonance}

\subsubsection{Slope Detection}

\begin{figure}[ht]
\begin{center}
   \includegraphics[width=0.8\columnwidth]{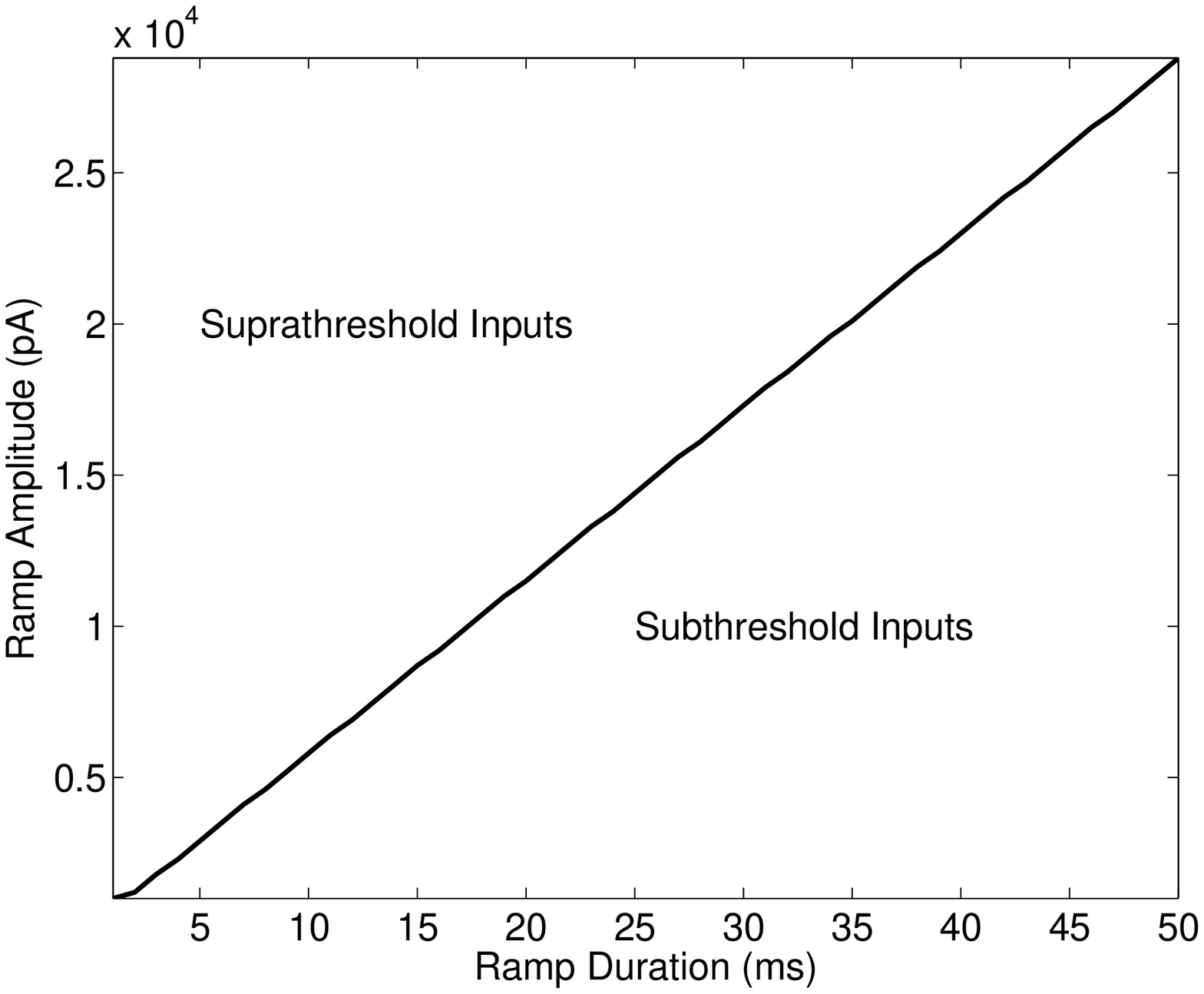}
  \caption{Slope-threshold of spiking response for a ramp stimulus, for the Type II model used to show slope-based stochastic resonance (SBSR). Ramp durations from $1$~ms to $50$~ms were simulated in $1$~ms intervals. For each of these simulations, the amplitude of the ramp input was increased by $100$~pA until threshold was reached. The threshold was defined by the first amplitude at which an action potential with $V_{\rm m}\geq 0$~mV was produced. The observed linear relation, except for non linear effects at small $A$ and $t$, shows that phasic neurons are best defined as slope (input derivative) detectors, rather than amplitude, detectors---see~\cite{Gai10}. }\label{f:thresh}
  \end{center}
\end{figure}

The behaviour of the Type-II (phasic) VCN neuron is best defined in terms of a slope threshold, pertaining to the rate of change of the input signal. This can be seen to manifest itself as the model only generating action potentials for ramp inputs above a constant ratio of amplitude to time window (see Figure~\ref{f:thresh}).
Classical stochastic resonance occurs when the amplitude of subthreshold signals is modified by stochastic noise that induces threshold crossings~\cite{McDonnell.PLOS09}. Since there is no such amplitude defined threshold for this VCN neuron, a  classical  stochastic resonance analysis was found to be insufficient~\cite{Gai10}. The resulting observation of stochastic facilitation effects  are instead related to the magnitude of the derivative of the input current; this constitutes SBSR.

 \subsubsection{Model Response to Low Frequency Sinusoidal Input With Channel Noise}

\begin{figure}[htbp]
\begin{center}
  \subfigure[]{\includegraphics[width=0.55\columnwidth]{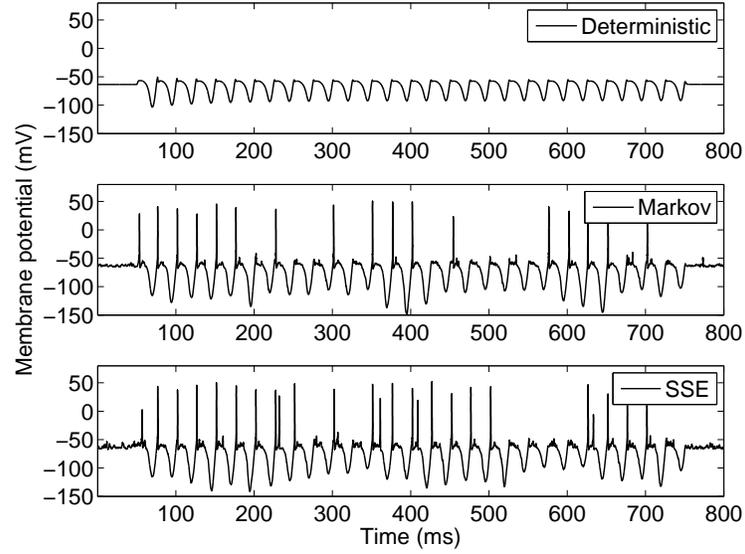}\label{f:SBSR}}
    \subfigure[]{\includegraphics[width=0.55\columnwidth]{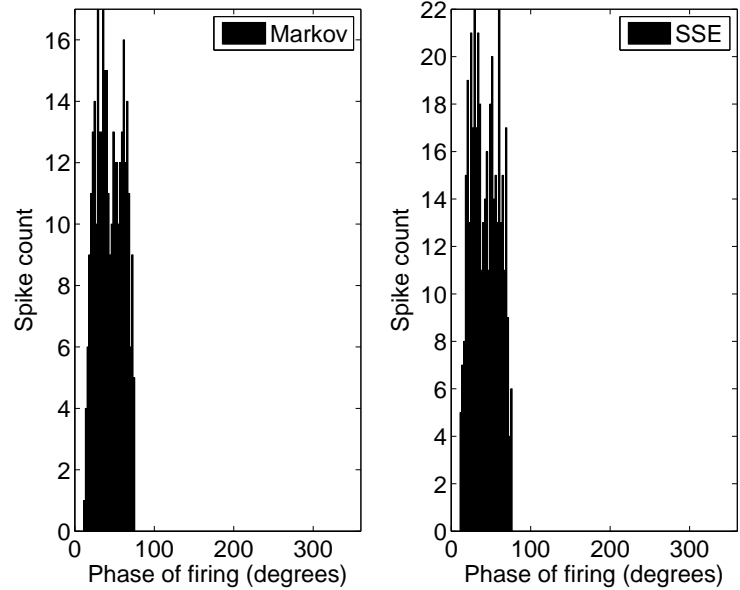}\label{f:SBSR_hist}}
  \caption{(a) Comparison of channel noise and deterministic models for a sinusoidal current input to the phasically firing neuron model. The traces show responses of a single neuron to a 0.55nA 40Hz sinusoidal input for the deterministic model and both the system size expansion (SSE) and Markov chain channel noise models. When spiking occurs, spike timing codes the frequency of the input, whereas there is no response from the deterministic model and thus the model shows SBSR when channel noise is the source of stochastic variability. Firing occurs prior to the maximum input amplitude, at the rising phase of the input signal; this occurs robustly for all sinusoidal inputs that elicit a response. (b) Peristimulus time histograms (PSTHs) for the Markov and SSE models; these figures show histograms of the phase at which the model produces an action potential, relative to the stimulating sinusoid, and confirm that firing always occurs prior to the maximum input current (which occurs at $90^o$)}\label{blah}
  \end{center}
\end{figure}

As the threshold for action potential generation is defined in terms of the rate of change of the input current, in the absence of noise the Type-II (phasic) VCN model neuron will not respond to slowly varying input currents, i.e., those with shallow slopes (see Figure~\ref{f:thresh}). Figure~\ref{f:SBSR} shows the model response to such a slope-referenced subthreshold sinusoidal input current. The deterministic model is seen to not respond to this signal, whereas with the inclusion of channel noise, for both the SDE and Markov models, a phasic response is observed.   This feature is noted to be robust across the range of inputs that illicit a response and further confirms slope based detection. When action potentials occur, they only occur during phases of the input stimulus corresponding to its region of maximum slope, rather than the phase corresponding to the maximum amplitude, as is shown in peristimulus-time histograms (PSTHs) in Figure~\ref{f:SBSR_hist}. Our channel noise models were not observed to respond to the falling phase of the sinusoidal input in any of the performed simulations, even when the number of channels is set unrealistically low---see Discussion.

\subsubsection{Model Response to Ramp Input With Channel Noise}

\begin{figure}[htbp]
\centering
   \includegraphics[width=1\columnwidth]{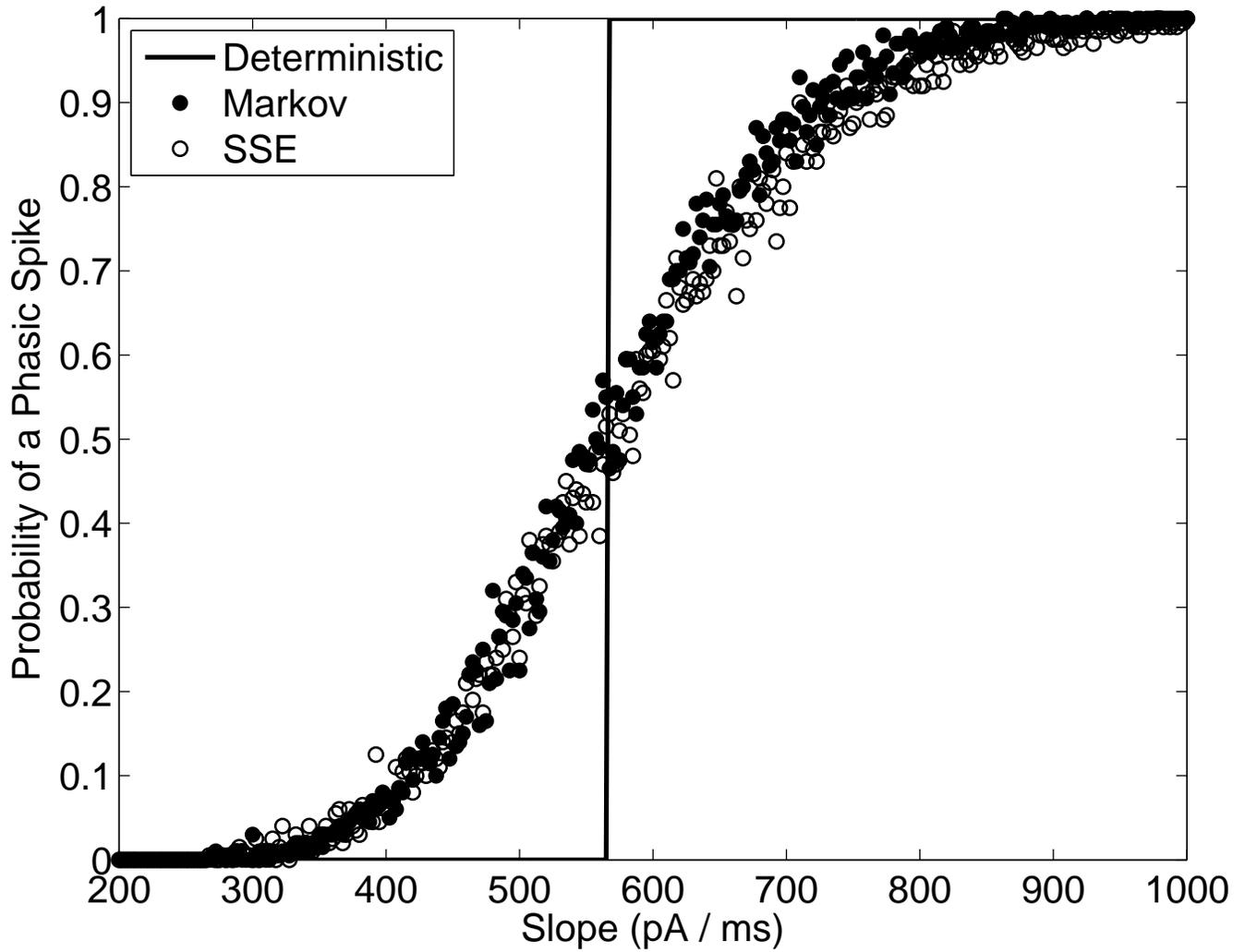}
  \caption{Estimated probability of initiating a phasic spike response in the Rothman and Manis model, for a ramp input current with a given input slope. For the system size expansion (SSE) and Markov chain ion channel noise models,  $N_{\rm a}=1000 $ channels for all ion channel types. These data were obtained by averaging over 200 repeats for each maximum amplitude (between 4 and 20 nA, in steps of 0.5 nA) of a fixed duration ramp (20ms). The model produces at most a single spike at the onset of the ramp. The  probability of a phasic spike  for a given slope value was estimated by summing the total number of repeats in which a phasic spike was produced, and dividing by the number of repeats (200).}
  \label{f:Ramp}
\end{figure}

 Figure \ref{f:Ramp} shows the estimated probability of eliciting a phasic spike for a given slope  of DC input. The deterministic case is seen to have a defined threshold at approximately $0.5$~nA.ms$^{-1}$. The channel noise models, however, show a region of input slopes that create a sigmoidal probability curve, meaning that for near-threshold inputs that this result is analogous to the findings in~\cite{Gai10} where  white noise was added to the input current.

\subsection{Inverse Stochastic Resonance}

Very small amplitude white noise has been shown to cause stochastic switching between a periodic limit cycle (indicating action potential generation) and a resting state in Hodgkin-Huxley  model neurons~\cite{Tuckwell,Gutkin,Tuckwell.09,Tuckwell.10}. This occurs for constant suprathreshold input currents that are close to the firing threshold for tonic firing in the deterministic version of the model. Moreover it was shown that there exists a noise amplitude that will minimize the average firing rate over many trials for a given input current. The occurrence of this minima, which is below the firing rate for the corresponding deterministic model, has led to the effect being labeled ISR~\cite{Tuckwell}. 

\subsubsection{ISR in Hodgkin-Huxley and Rothman-Manis Type I-II models}

\begin{figure}[htbp]
\centering
   \subfigure[~Hodgkin-Huxley model]{\includegraphics[width=0.45\columnwidth]{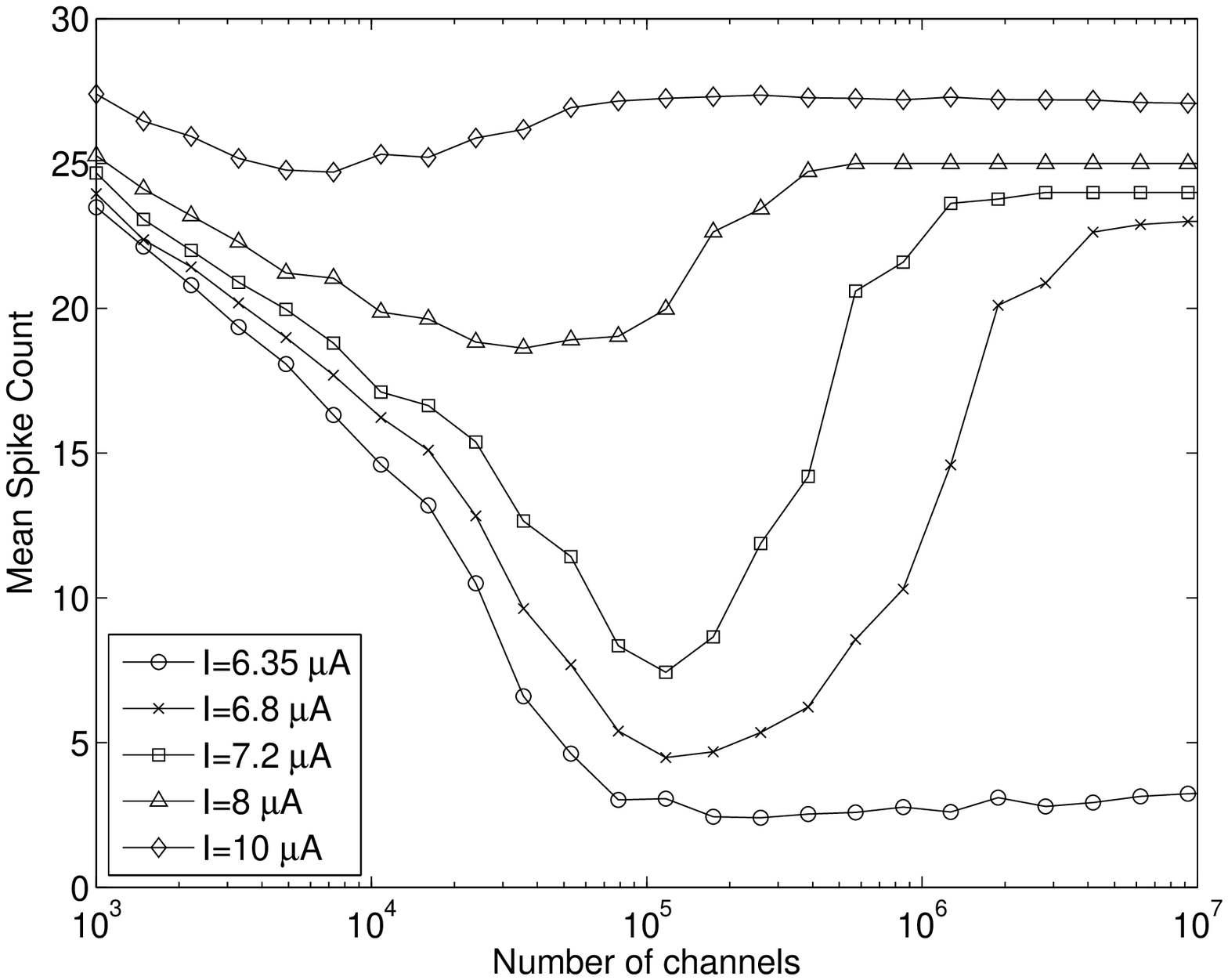} \label{f:HHISR}}
    \subfigure[~Rothman-Manis Type I-II]{\includegraphics[width=0.45\columnwidth]{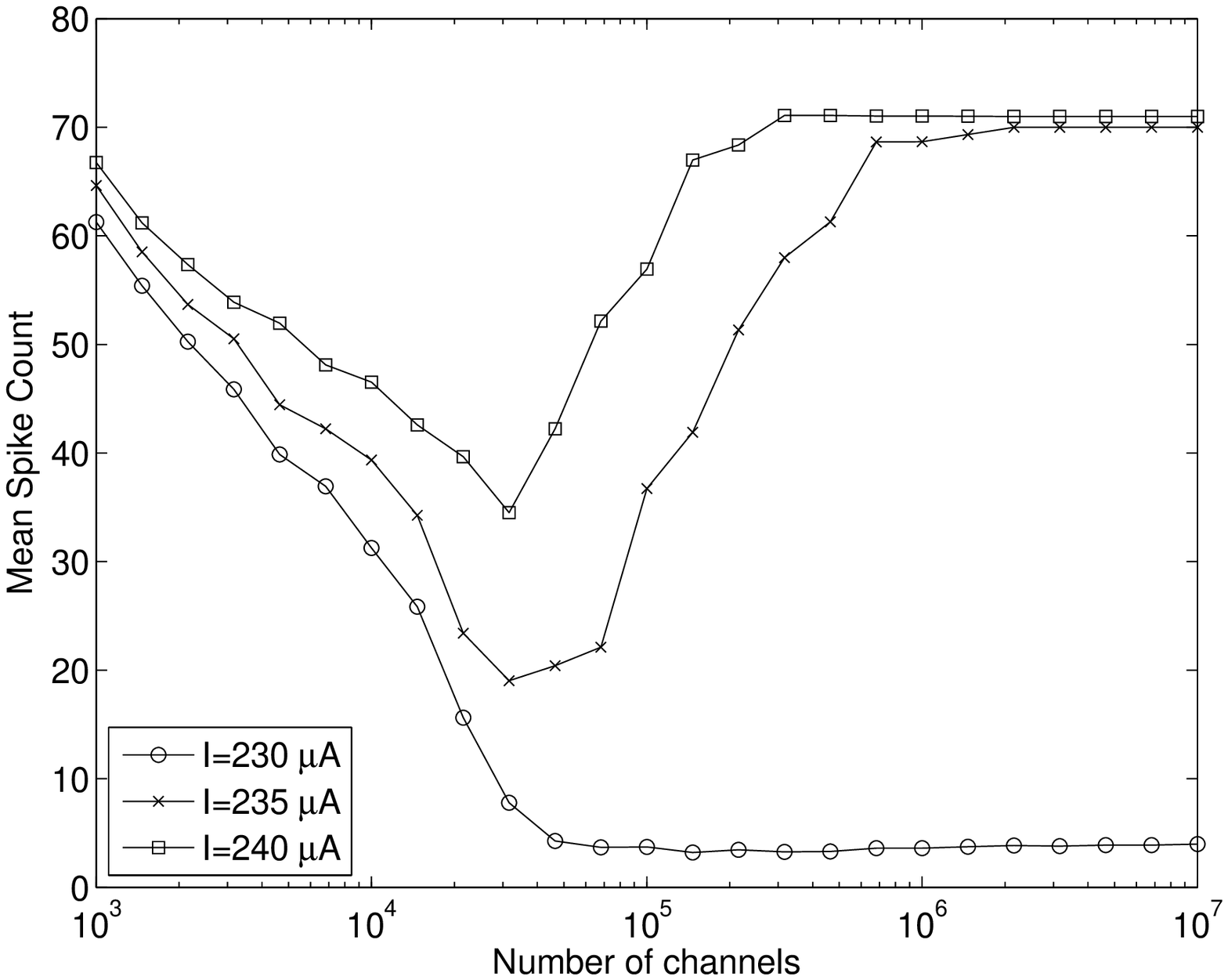}  \label{f:RothISR}}
  \caption{Inverse Stochastic Resonance in the (a) Hodgkin-Huxley and  (b) Rothman and Manis~\cite{RothmanManis2003b} Type I-II VCN neuron models, using channel noise as the source of stochasticity. The traces show the mean number of spikes in a 400~ms time window (ensemble averaged over 100 repeats) from the onset of the input current, versus the number of channels, $N_{\rm a}$, where $N_{\rm a}$ is the same for all ion channel types. Note the effective noise amplitude decreases from left to right, i.e. noise variance decreases as the number of channels increases. The deterministic threshold for tonic firing in the Hodgkin-Huxley model is at $I\simeq 6.45~\mu$A. In the deterministic model, for $I=6.8$ there were 23 spikes, for $I=7.2$ there were 24 spikes, for $I=8$ there were 25 spikes and for $I=10$ there were 27 spikes. The threshold for tonic firing in the deterministic Type I-II model is $I\simeq 234~\mu$A. For $I= 235$ there were 70 spikes in the deterministic model in the 400~ms time window. For $I=240$, there were 71 spikes in the 400~ms time window. For the case of $I=230$, approximately 10\% of repeats for $N_{\rm a}>10000$ exhibited sustained tonic firing for the entire 400~ms rather than quiescence following a short onset burst. These points were removed for the purposes of this figure.}
\end{figure}

Here, our results in Figure \ref{f:HHISR} show that modifying the Hodgkin-Huxley model used by~\cite{Tuckwell,Gutkin,Tuckwell.09,Tuckwell.10} to include  channel noise, instead of extrinsic injected current noise, also lead to ISR.  We have extended upon this to show that the tonically firing Type I-II (Figure \ref{f:RothISR}) version of the Rothman-Manis model~\cite{RothmanManis2003b} also exhibits this effect. We note that the number of channels at which the ISR minimum firing rate occurs is between $10^4$ and $10^5$ channels. However, the firing rate of the deterministic model can be seen to decrease for channel numbers of the order of $10^6$, and therefore  stochastic effects due to ion channel noise are present near threshold for large $N_{\rm a}$. 

\begin{figure}[htbp]
\centering
 \subfigure[~Membrane potential]{ \includegraphics[width=0.45\columnwidth]{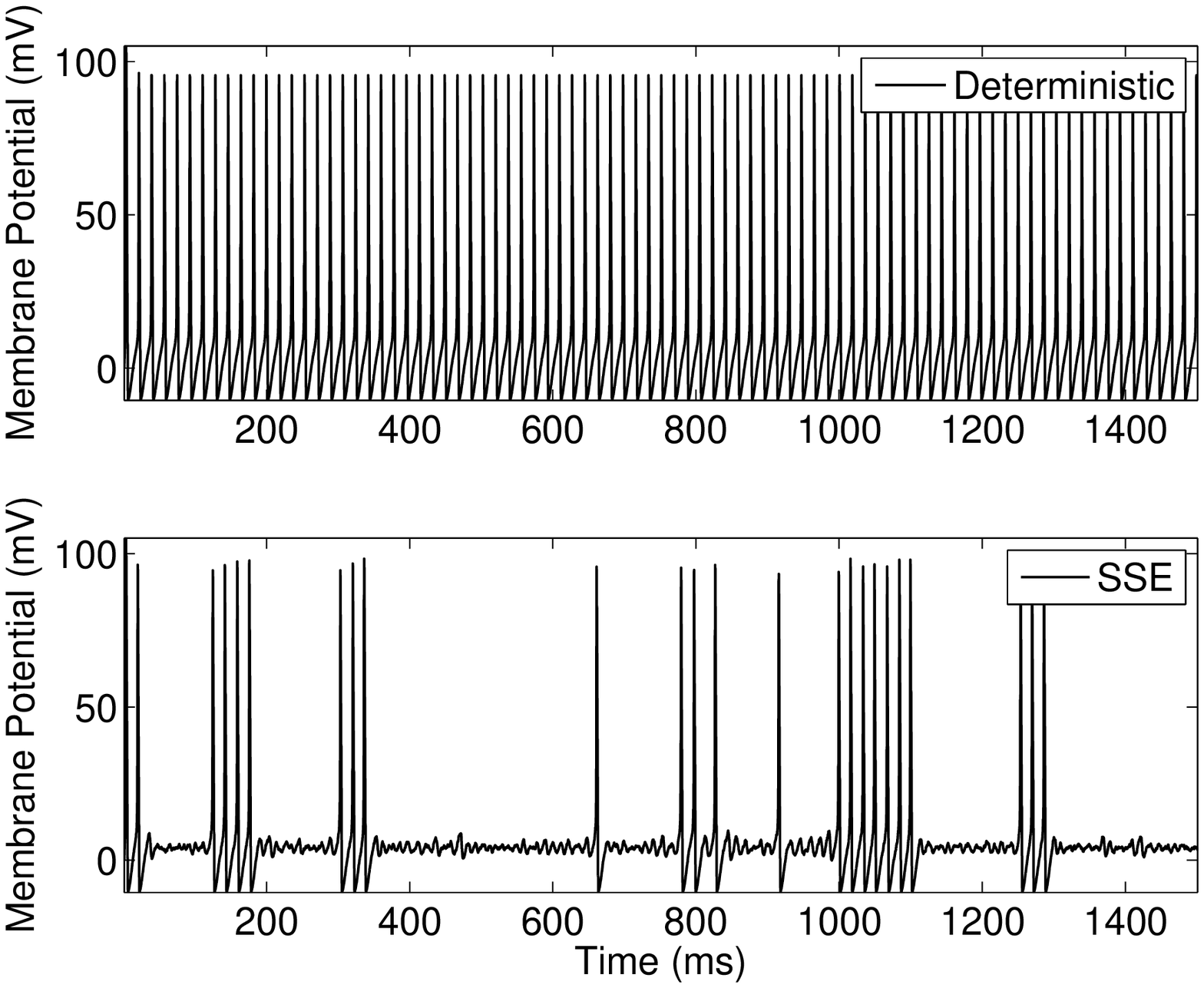} \label{f:HHtrace}}
  \subfigure[~ISI histogram for SSE]{ \includegraphics[width=0.45\columnwidth]{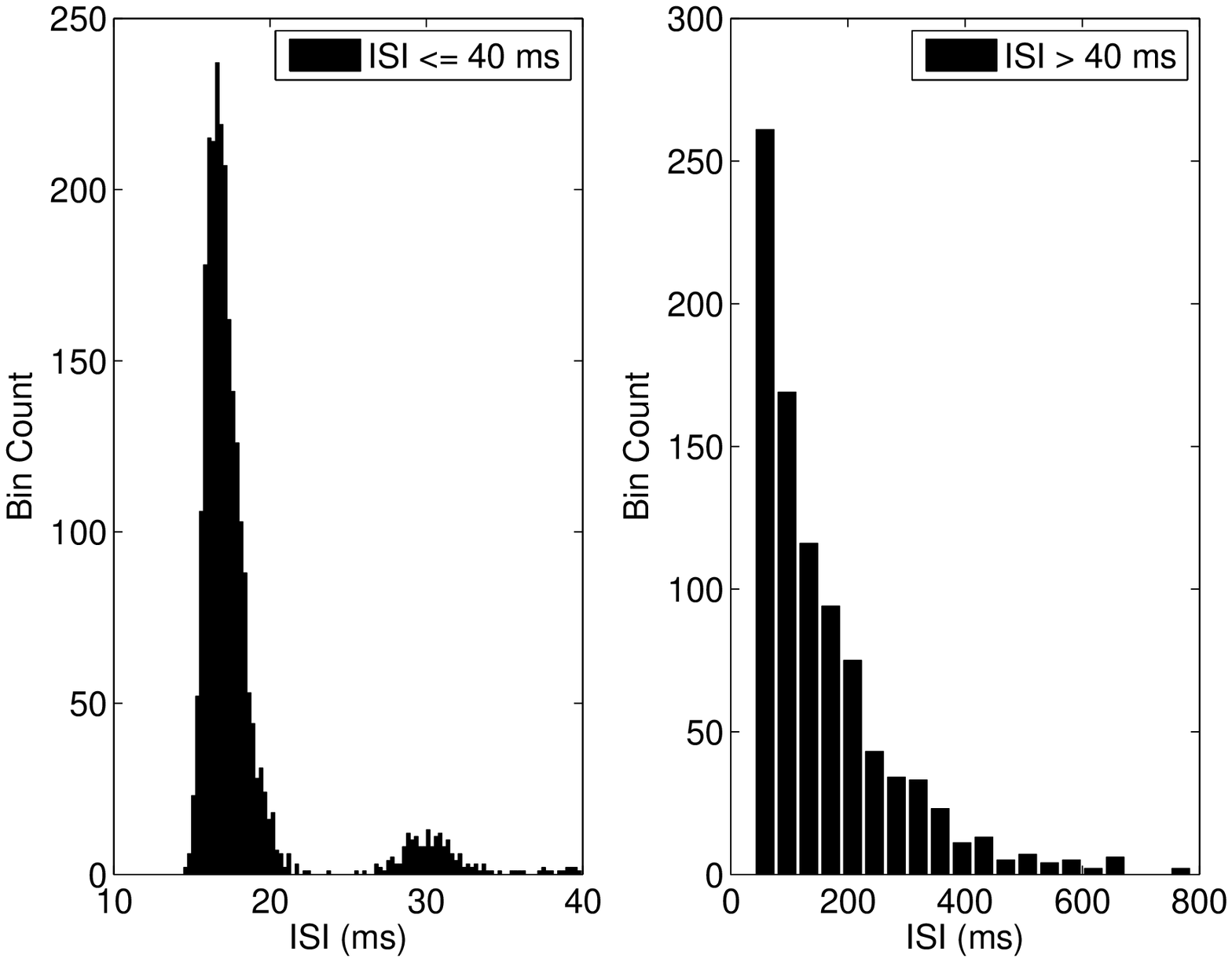} \label{f:HHtrace_hist}}
  \caption{(a) Representative membrane potential traces for the Hodgkin-Huxley model for both the deterministic (top) and channel noise (SSE) models, with $N_{\rm a}$=45000 for all ion channel types and a suprathreshold DC input current = $6.8~\mu$A. SSE indicates  results obtained by simulating ion channel noise using the system size expansion method. (b) ISI histograms for the SSE model, obtained from an ensemble of 100 repeats of 2 second duration simulations.  The left subpanel in (b)  indicates that intra-burst ISIs have low variance, while the right sub panels indicate a large variance in the interval between bursts.  The bimodal characteristic of the inter-spike-intervals is consistent with results in~\cite{Rowat.07}. Note that the steady-state ISI for the deterministic model with suprathreshold stimulation was approximately 17.8 ms.}

\end{figure}

We discuss the mechanism of ISR in detail in Section~\ref{S:Discuss_ISR}, where we refer to the following results.  The presence of noise  can cause switching back and forth between two co-existing stable states---a resting state  (R) and a limit cycle (L). We label the probabilities of a transition as $P_{R \rightarrow L}$ and $P_{L \rightarrow R}$ respectively. As can be seen in Figure~\ref{f:HHtrace},  switching between the limit cycle and resting state may occur frequently for the Hodgkin-Huxley model with channel noise, as is examined in detail for the case of external current noise sources in~\cite{Gutkin,Tuckwell} for this model. For the input current and number of channels simulated in Figure~\ref{f:HHtrace}, the Hodgkin-Huxley model is seen to dwell longer in the resting state than within the limit cycle. This is also evident in Figure~\ref{f:HHtrace_hist}, where the ISI histogram has a long tail showing many ISIs greater than the frequency of the deterministic limit cycle, thus indicating that $P_{L \rightarrow R} > P_{R \rightarrow L}$.

\begin{figure}[htbp]
\centering
\subfigure[~Subthreshold input]{ \includegraphics[width=0.4\columnwidth]{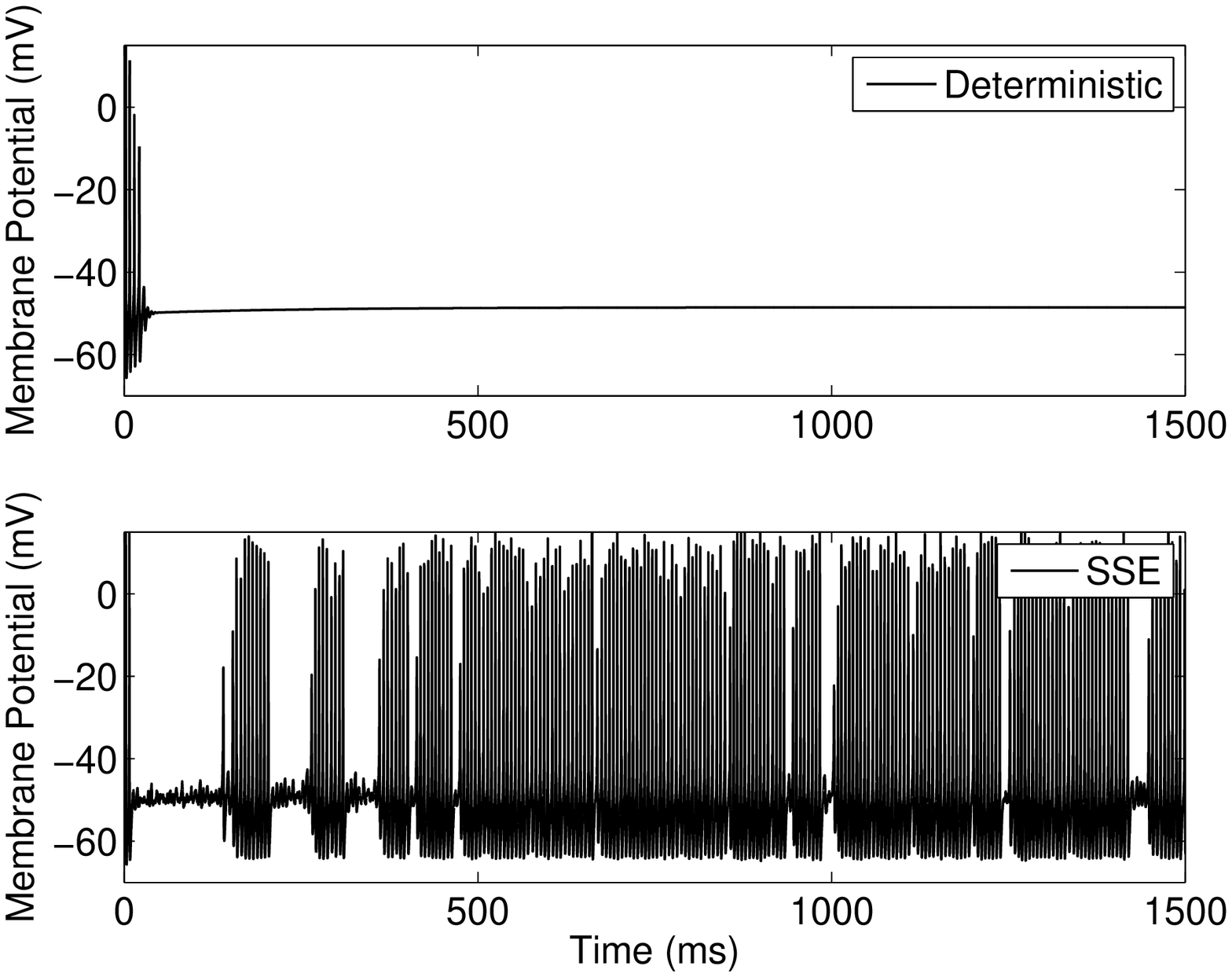}\label{f:ItoIIsub}}
\subfigure[~Suprathreshold input]{ \includegraphics[width=0.4\columnwidth]{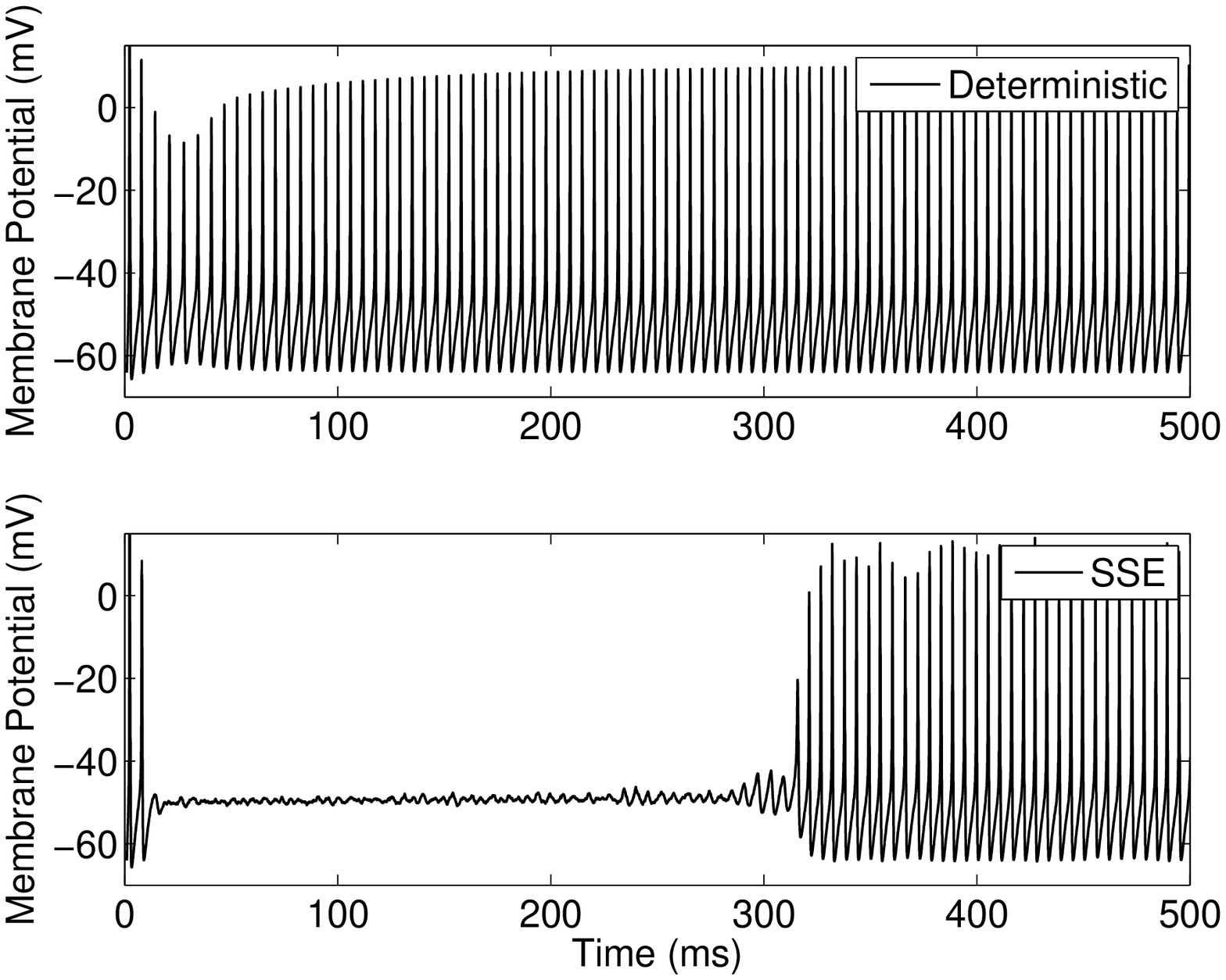}\label{f:ItoIIsup}}

\subfigure[~Subthreshold input, 10000 channels, SSE]{ \includegraphics[width=0.4\columnwidth]{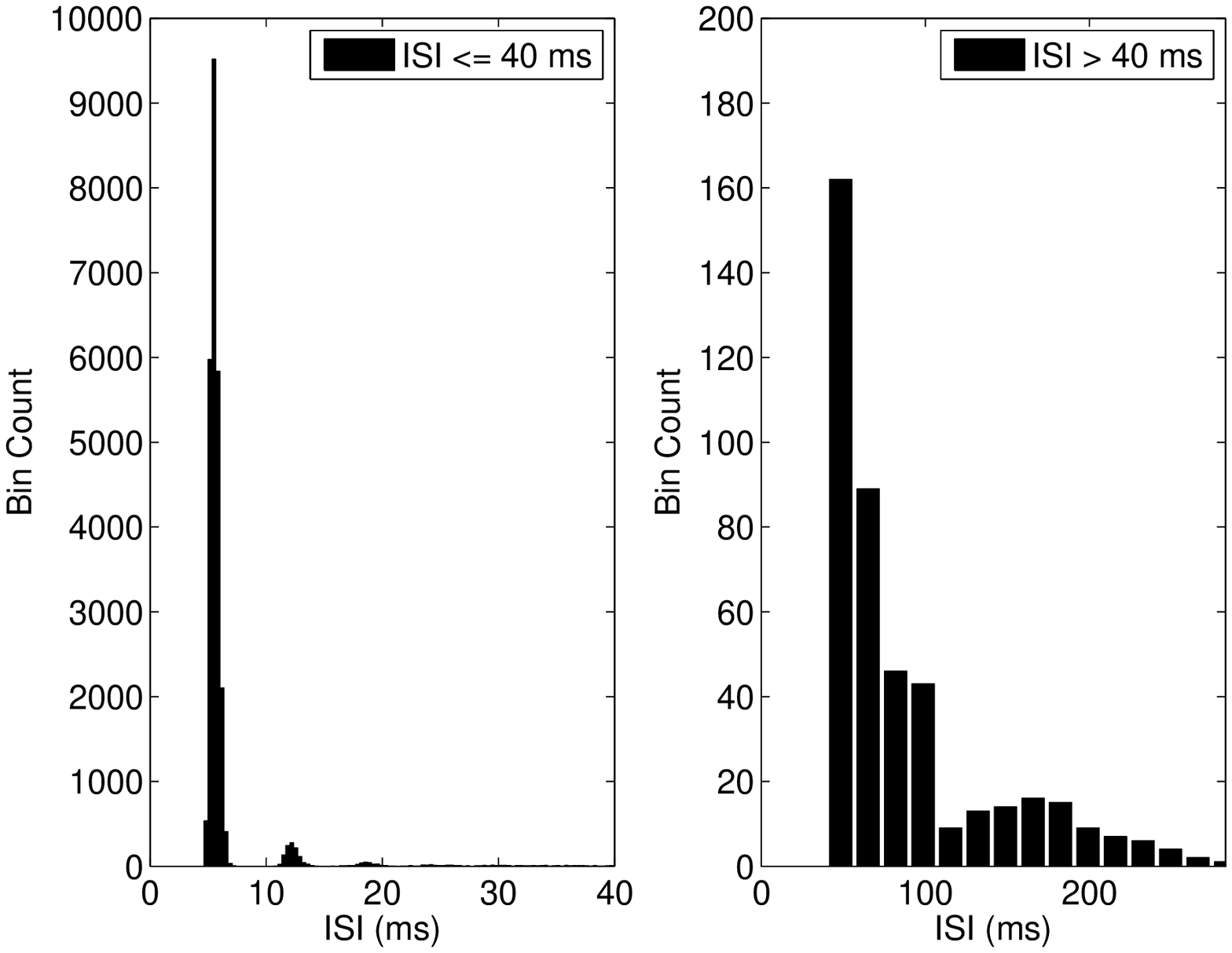}\label{f:ItoIIsub_hist}}
\subfigure[~Suprathreshold input, 50000, channels, SSE]{ \includegraphics[width=0.4\columnwidth]{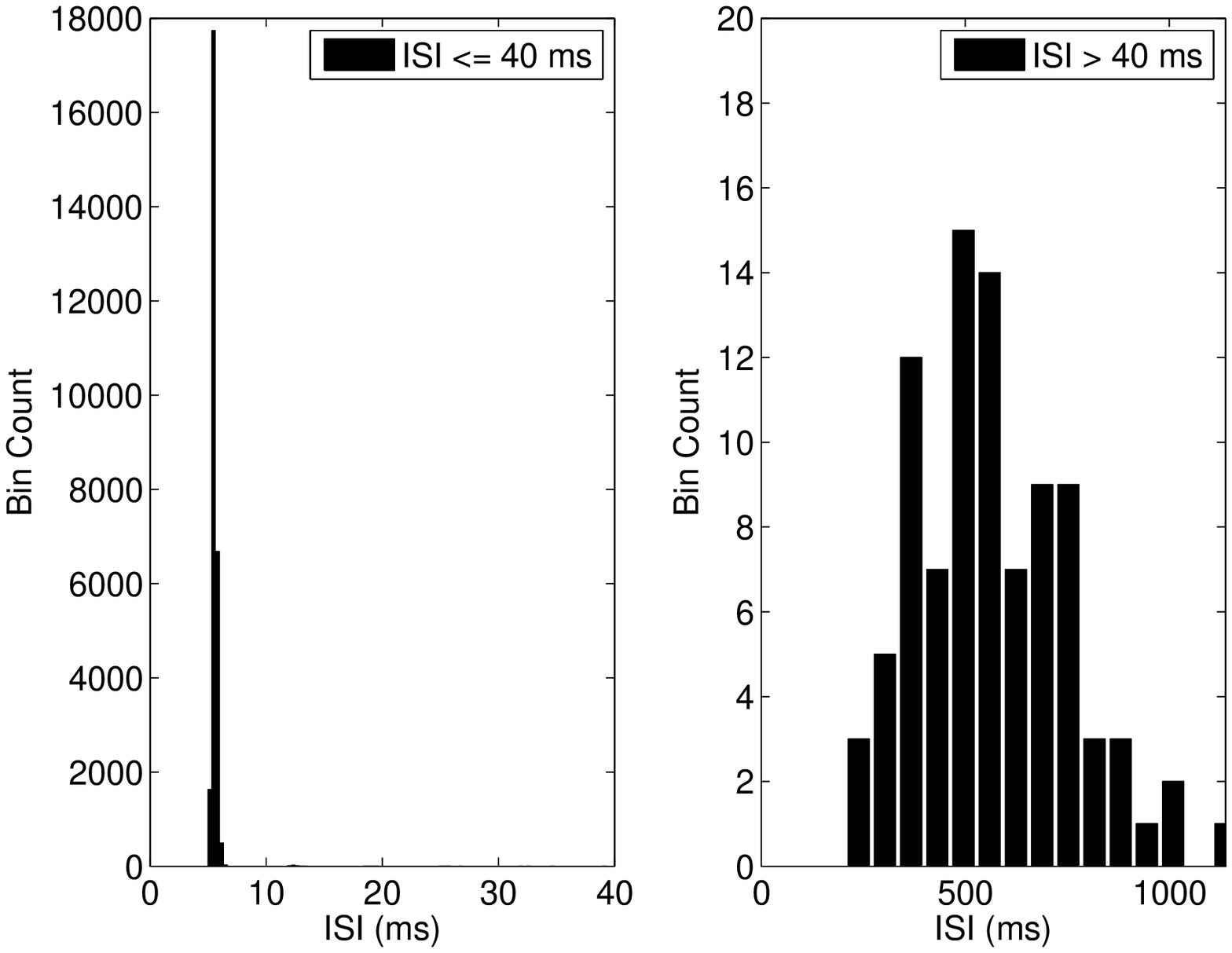}\label{f:ItoIIsup_hist}}
  \caption{Representative membrane potential traces, and ISI histograms, for the Type I-II Rothman-Manis model for both the deterministic and channel noise (SSE) models, with two different numbers of channels. (a) Membrane potential for  a subthreshold DC input current of $I= 234~\mu$A, starting at $t=0$, for the deterministic model (top) and the SSE model with $N_{\rm a}=10000$ for all ion channel types.  (b) Membrane potential  for a suprathreshold DC input current of $I= 236~\mu$A starting at $t=0$, for the deterministic model (top) and the SSE model with $N_{\rm a}=50000$ for all ion channel types. SSE indicates  results obtained by simulation ion channel noise using the system size expansion method. Note that the sub- and supra-threshold behaviour with the SSE model are very similar for the same number of channels, and so we have used two different channel numbers here to highlight different noise levels. (c) ISI histograms for subthreshold stimulation using the SSE model. (d) ISI histograms for suprathreshold stimulation using the SSE model. The left subpanels in (c) and (d) indicate that intra-burst ISIs have low variance, while the right sub panels indicate a large variance in the interval between bursts. Note that the steady-state ISI for the deterministic model with suprathreshold stimulation was approximately 5.5 ms.}\label{f:I_11}
\end{figure}

Within the deterministic Type I-II model configuration, introducing a constant input current just below the threshold for tonic firing produces an initial phasic burst of action potentials. Near the end of this burst the spike amplitude decreases until only a subthreshold oscillation is observed (Figure~\ref{f:ItoIIsub}, top). If however a suprathreshold input for tonic firing is applied, the model will continue to fire after this initial bursting behavior and the spike amplitude will gradually grow back to a steady height (Figure \ref{f:ItoIIsup}, top).

For the stochastic model, the possibility of state switching causes the behavior of the model to be similar for suprathreshold and subthreshold inputs that are close to threshold. Examination of the behavior of the model in the near-threshold region gives insight into the mechanisms that produce the ISR effect. At large numbers of ion channels (in the order of $10^6$) the noise amplitude is low and once the tonic firing state is entered after the initial phasic burst, entry back into the resting state was not observed for the longest simulations performed in this study (1 minute of simulated activity).  This indicates that $P_{L \rightarrow R}\approx 0$ for large channel numbers during the tonically firing limit cycle. For fewer channels, a response such as that shown in Figure~\ref{f:ItoIIsup}(bottom) is obtained; once the stochastic model enters the rest state, it will  eventually return to the limit cycle.

However, this change from the limit cycle to the resting state may occur ($P_{L \rightarrow R}> 0$) in the region of lower amplitude spikes which occurs approximately $100$~ms after onset of the input current (Figure~\ref{f:ItoIIsup}, top). In this model configuration $P_{R \rightarrow L}$ is lower for near-threshold inputs at larger values of $N_{\rm a}$ than in other models considered here. Again, this may be observed in Figure~\ref{f:ItoIIsup}(bottom) where the delay after the initial spikes before the limit cycle is entered may be greater than one second for channel numbers corresponding to the minima in Figure~\ref{f:RothISR}. This behavior may also be observed for subthreshold inputs at large channel numbers ($N_{\rm a}\approx 10^5$) but entry into the limit cycle occurs less frequently than for suprathreshold inputs.

At channel numbers in the order of $10^4$, due to the larger noise amplitude, switching between states may be observed within the time course of simulations as in Figure \ref{f:ItoIIsub}(bottom). This rapid switching between states is also observed for suprathreshold inputs that are close to threshold with low channel numbers. This overall behavior leads to the minimum in average firing rate observed in Figure~\ref{f:RothISR}, characteristic of the ISR effect being strongly observed.

%%%

\subsubsection{Rothman-Manis Type I-c model}

\begin{figure}[htbp]
\centering
 \subfigure[~Membrane potentials]{\includegraphics[width=0.45\columnwidth]{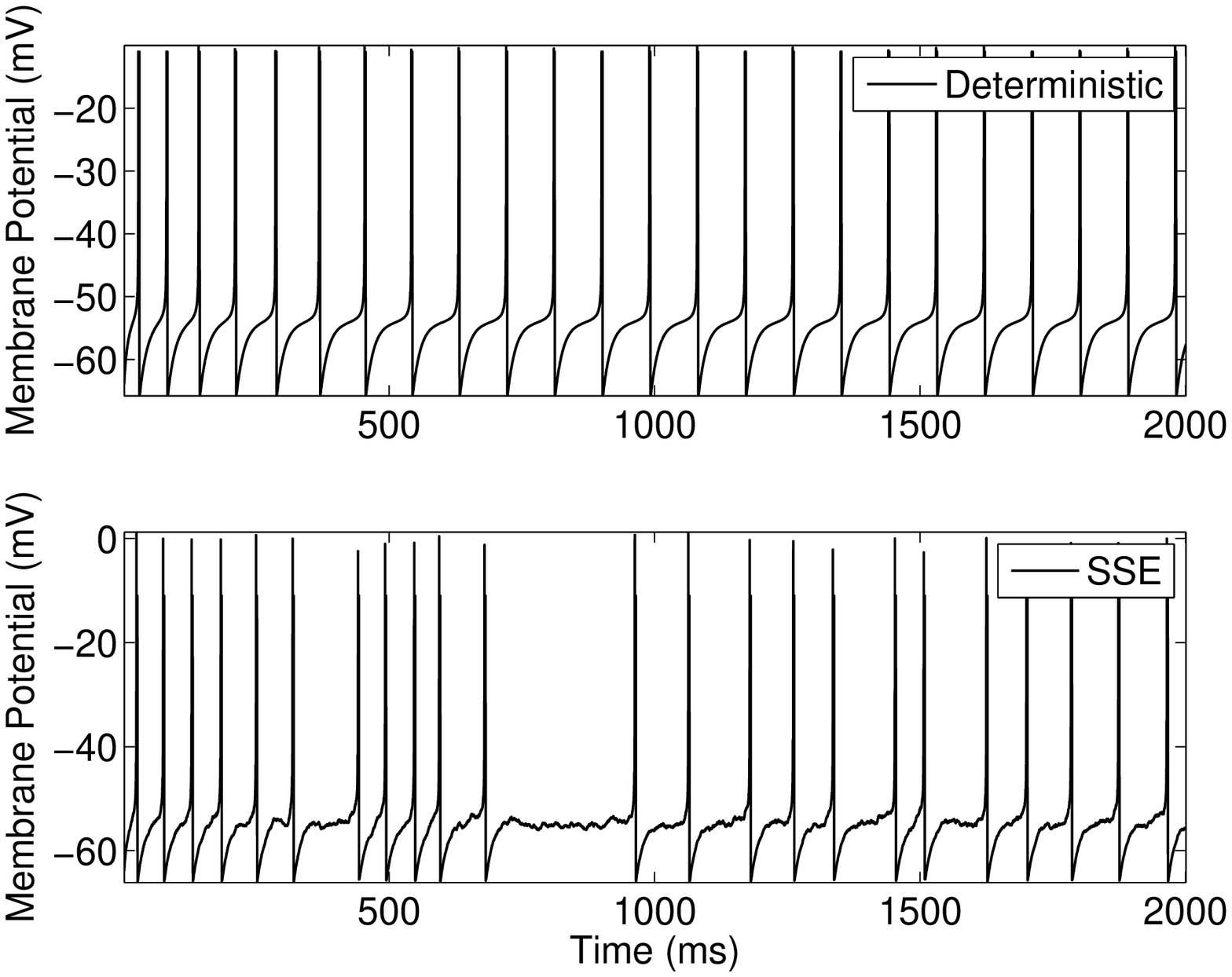}  \label{f:ItoCtrace}}
 \subfigure[~ISI histograms]{\includegraphics[width=0.45\columnwidth]{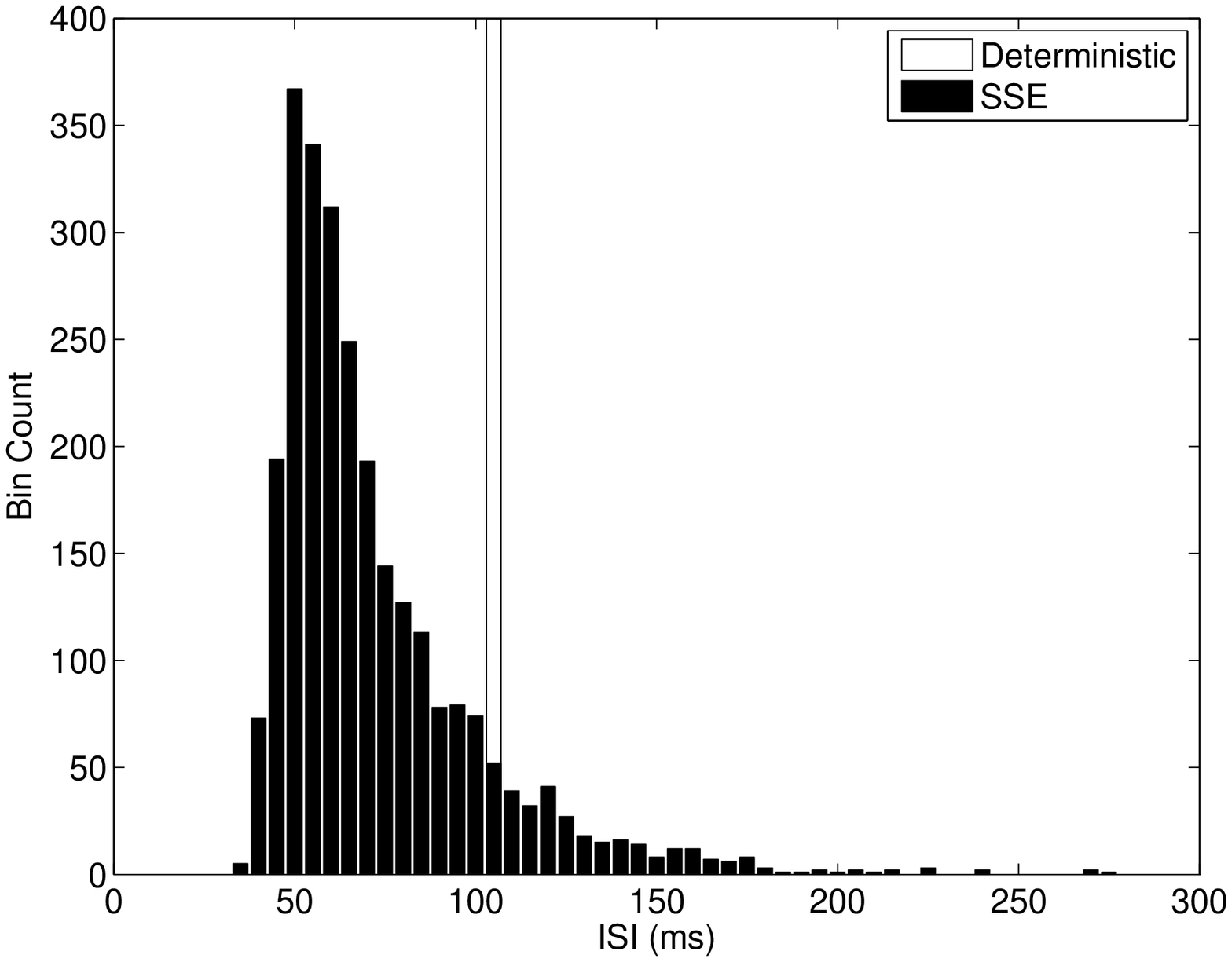}  \label{f:ItoChist}}
  \caption{(a) Membrane potential of the Type I-c Rothman-Manis model for both the deterministic and channel noise (SSE) models. The traces were obtained for  $N_{\rm a}$=50000 for all ion channel types and a DC input current $I = 25.75~\mu$A (just above threshold for tonic spiking), starting at t=0. (b) Histogram of the inter-spike-intervals (ISI)  for a DC input current $I = 25.75~\mu$A, obtained  from an ensemble of 100 repeats of 2 second duration simulations, with  $N_{\rm a}=50000$ for all ion channel types. SSE indicates  results obtained by simulation ion channel noise using the system size expansion method. The data shows that channel noise increases the average spike rate relative to the deterministic model, and unlike the Type I-II model channel noise does not induce spike bursting behaviour.}

\end{figure}

Finally, we considered the Rothman-Manis Type I-c model~\cite{RothmanManis}; the primary difference between this model and the Type I-II model is that the former lacks a low threshold potassium current.   We found that unlike the Type I-II model, that the Type I-c model does not exhibit ISR, and instead shows only an increased mean spike rate as the number of channels decreases, for suprathreshold input current (Figure~\ref{f:ItoCtrace}). The reason why this model does not show ISR is discussed in Section~\ref{S:Discuss}.

We also found that the Type I-c model shows a much stronger increase in firing rate to suprathreshold inputs in the near-threshold region than both the Hodgkin-Huxley model and Type I-II model. As shown in Figure~\ref{f:ItoChist}, inclusion of channel noise in the Type I-c model leads to  a  significant peak in the Inter-Spike Interval (ISI) at intervals shorter than the deterministic ISI value. In contrast, the other models considered here do not show significant peaks in the ISI for values shorter than the deterministic value (Figures~\ref{f:HHtrace_hist} and~\ref{f:I_11}). 

\section{Discussion}\label{S:Discussion}

\subsection{Slope-based stochastic resonance}

We have demonstrated, using both sinusoidal and ramp inputs, that slope-based stochastic resonance can be observed due to channel noise. In the case of sinusoidal inputs, we found that channel noise induced spiking in response to low frequency inputs occurs only during the rising phase. However~\cite{Gai09} also reported firing during the falling phase of the sinusoidal input, when using injected current noise.  This confirms the comment of~\cite{Gai09}, that unrealistically high noise levels were used to generate action potentials on the falling phase of the input.

In the case of a ramp input current, SBSR was demonstrated by production of a sigmoidal  increase of firing rate with increasing slope. It has been shown previously that such a sigmoidal shape in the probability of firing may be exploited for increased information transmission in a parallel {\em population} of stochastic Hodgkin-Huxley models~\cite{Ashida}. This previous demonstration by~\cite{Ashida} of a form of stochastic facilitation, known as suprathreshold stochastic resonance~\cite{Stocks.Mar2000,McDonnell.07}, due to channel noise is consistent with the central conclusions of the current paper.

\subsection{The stochastic dynamics that causes ISR}~\label{S:Discuss_ISR}

We have demonstrated that ISR can occur due to channel noise in the Hodgkin-Huxley and Rothman-Manis Type I-II models. As for the case of injected current noise~\cite{Tuckwell,Gutkin,Tuckwell.09,Tuckwell.10}, the effect is observed due to stochastic switching between a limit cycle and a fixed point. The minima in the ISR curve occurs when the noise level is such that the probability of initial switch from limit-cycle to fixed point, followed by a long dwell in the fixed point, is highest.

Both the Hodgkin-Huxley and Rothman-Manis models exhibit Class-II excitability~\cite{Izhikevich}, which means that the deterministic forms of the models exhibit high tonic firing rates above some threshold value of constant injected current, $I$, and no spiking below the threshold, such that spike rates discontinuously jump from zero to a large value as $I$ increases. This behavior can be attributed to the existence of  a subcritical Andronov-Hopf bifurcation~\cite{Izhikevich}. As noted by~\cite{Tuckwell.09}, the combination of this dynamics with noise means the model can stochastically switch between the resulting stable limit-cycle and fixed point, thus leading to bursting behavior that causes ISR. Such noise-induced bursting is consistent with the results of a previous detailed study of channel-noise induced bursting in the Hodgkin-Huxley model~\cite{Rowat.07}.  

On the other hand, we did not observe ISR in the Rothman-Manis Type I-C model. The Type I-C model has a much lower threshold to tonic spiking in response to constant current, I, in comparison with the Type I-II model, and as mentioned lacks a low threshold potassium current. This is sufficient to change the dynamics of the system such that the deterministic form of the model exhibits Class-I excitability, which means that as the input current $I$ is brought above threshold, spike-rates increase continuously from zero. Such behavior is attributable to the existence of a saddle-node-on-invariant-circle bifurcation, which means a stable fixed point disappears for above-threshold current levels~\cite{Izhikevich}. Thus, the system acts like an integrator (as noted by~\cite{RothmanManis}) and there is no scope for bistable switching from a limit-cycle to a rest-state, and hence no opportunity for ISR to be observed.

\subsection{Effects of stochastic ion channels on spiking dynamics}~\label{S:Discuss}

The extent of the stochastic behavior observed in all models due to channel noise may be qualitatively understood by considering the case where the model neuron's membrane potential is close to causing action potentials to occur. In this regime, fluctuations in all membrane currents can contribute to whether or not an action potential occurs. The size of the fluctuations in each current are determined by variance in the number of open channels, $\mathcal{N}_{\rm a,o}(t)$, the associated  peak conductance, and the difference between the membrane potential and the reversal potential. 

Since the variance of an activation variable is high when $\mathcal{N}_{\rm a,o}(t)$ is low~\cite{Schneidman,Goldwynwhat}, if such a current has a large peak conductance, the path variance of the membrane potential may also be significantly affected in this case. 

For the models with Class-II excitability in which ISR occurs, the sodium channel has the largest peak conductance and a value of $G_{\rm Na}(t)$ that is close to zero when the membrane potential is close to threshold. Thus, the occurrence of large $P_{R \rightarrow L}$ and $P_{L \rightarrow R}$ that results in rapid state switching for low channel numbers can be attributed primarily to noise in the sodium channels. Indeed, we found that removing the stochastic noise from only the sodium channels meant that much lower channel numbers in remaining channel types was required to see state switching.

\subsection{Simulating large numbers of channels}

The small noise amplitude required to increase firing rates above the minimum as noise decreases in ISR  corresponds to a large number of channels, showing that stochastic effects due to ion channel noise are present near threshold for large $N_{\rm a}$. This is contradictory to the frequently used assumption that for large $N_{\rm a}$ the fluctuations due to channel noise may be neglected, but is consistent with other previous studies of the influence of channel noise~\cite{Schneidman,Shuai.05,Faisal.07,Goldwynwhat}. 

Our simulations took advantage of the fact that the run time of the SSE noise model does not scale with $N_{\rm a}$, as the Markov model does. Therefore large numbers of ion channels were able to be efficiently simulated, allowing this effect to be elucidated.

\subsection{Estimating realistic numbers of channels for each ion channel type}

As mentioned above, our results assume that all channel types have the same number of channels. This is a simplification made for the purposes of assessing whether channel noise can impact on spiking in a similar manner to current noise. Having established that it can, it is of interest to consider the influence of numbers of channels that vary realistically for each channel type. 

Various methods can be used to estimate total numbers of channels, e.g.~see~\cite{Buchholtz.02}.  Here, however, we make use of existing data on the conductance of {\em individual channels} from
~\cite{Gutman.05,Hofmann.05,Catterall.05} and then divide the total conductance for each channel type in our models by this value to obtain estimates.  Since total conductance is temperature dependent, and the temperature at which the individual channel conductances were measured is not available, we use the total conductances used in~\cite{RothmanManis2003b}.   This procedure is facilitated by discussion in~\cite{RothmanManis} of the molecular identities of each channel type in the Rothman-Manis model.  Our results are tabulated in Table 2, including references to papers used to obtain the single channel conductances.

The results in Table 2 are consistent with the following analysis. Using the neuronal diameter for the VCN neuron 21~$\mu m^{2}$~\cite{RothmanManis2003b} and assuming channel densities are about 18 per~$\mu {\rm m}^{2}$, for potassium channel and about 60 per~$\mu {\rm m}^{2}$ for sodium channels  as in Hodgkin-Huxley type models~\cite{Schneidman}, the number of channels in a spherical soma is estimated at being about 25000 potassium channels and about 83000 sodium channels. 

Our  first order estimations place the estimated number of channels \textit{in vivo} in a region where stochastic effects are strongly observed in these simulations.

\subsection{Conclusions}\label{S:Conclusions}

When channel noise is accounted for in model neurons, the response for inputs close to a firing threshold (whether an amplitude or slope threshold) may nontrivially differ from the deterministic model. Here we have shown that with the inclusion of intrinsic channel noise in a model of a VCN neuron that stochastic effects significantly alter the close to threshold behavior for both a phasic firing and a tonic firing configuration of the model, showing the effects of SBSR and ISR respectively. For the effect of SBSR both the Markov chain model and the SSE model were used to show the effect.  Thus our study suggests that it would be interesting  in future work to carry out detailed analysis of whether coding and processing of acoustic stimuli in the auditory brainstem may benefit from channel noise.

The assumption that for a large number of ion channels the fluctuations due to individual ion channels may be disregarded was found to be incorrect for some cases. We emphasise that the path of the membrane potential in phase space may have a large variance due to intrinsic channel noise for neurons in which the current making the strongest contributions to the membrane potential at threshold has low occupancy in the conducting (open) state. In agreement with this, stochastic effects were found to strongly influence behavior of the tonically firing models considered here at channel numbers in the range of those found in biological neurons. The different forms of stochastic behavior displayed by these tonic models were able to be qualitatively explained using considerations of the near-threshold value of $\mathcal{N}_{\rm a,o}(t)$, the associated membrane conductances and the near-threshold behavior of the model. These considerations when applied to the phasically firing Type II model predict that for a stochastic effect to be seen, the number of channels must be quite low to cause variance in dominating low threshold current, as indeed is the case.

In order to perform simulations for large numbers of individual ion channels the SSE model~\cite{FoxLu,Goldwyn} was used. Future work may include comparing the results of these simulations with other stochastic differential equation approximations to the Markov chain description, such as those found in~\cite{Schmandt,Dangerfield,Orio.12,Huang.13}. It will also be useful to consider stochastic facilitation effects due to channel noise in more detailed multi compartmental models.

\appendix

\section{Kinetics equations for the Rothman-Manis model}

This appendix lists the equations describing the kinetics of the 6 activation variables, $m,h,n,p,w,z$ and $r$,  used in the Rothman-Manis model, as originally stated in~\cite{RothmanManis2003b} but adjusted to account for temperature as in~\cite{Gai09}. Note that all activation variables and the membrane potential, $V$, are time-dependent variables, but to make the notation compact this has not been shown in the equations.

\begin{equation}
    m_{\infty}(V)= \frac{1}{1 + e^{-(V+38)/7}},\\
\end{equation}
\begin{equation}
    \tau_{m}(V)=\frac{10}{15e^{(V+60)/18}+108e^{-(V+60)/25}} +\frac{1}{75},
\end{equation}
\begin{equation}
    h_{\infty}(V)= \frac{1}{1 + e^{(V+65)/6}},\\
\end{equation}
\begin{equation}
    \tau_{h}(V)=\frac{100}{21e^{(V+60)/11}+30e^{-(V+60)/25}} +0.2,
\end{equation}
\begin{equation}
    n_{\infty}(V)= [1 + e^{-(V+15)/5}]^{-1/2},\\
\end{equation}
\begin{equation}
    \tau_{n}(V)=\frac{100}{33e^{(V+60)/24}+63e^{-(V+60)/23}} +\frac{7}{30},
\end{equation}
\begin{equation}
    p_{\infty}(V)= \frac{1}{1 + e^{-(v+23)/6}},\\
\end{equation}
\begin{equation}
    \tau_{p}(V)=\frac{100}{12e^{(V+60)/32}+15e^{-(V+60)/22}} +\frac{5}{3},
\end{equation}
\begin{equation}
    w_{\infty}(V)= [1 + e^{-(V+48)/6}]^{-1/4},\\
\end{equation}
\begin{equation}
    \tau_{w}(V)=\frac{100}{18e^{(V+60)/6}+48e^{-(V+60)/45}} +0.5,
\end{equation}
\begin{equation}
    z_{\infty}(V)= \frac{1}{2 + 2e^{-(v+71)/10}}+0.5,\\
\end{equation}
\begin{equation}
    \tau_{z}(V)=\frac{1000}{3e^{(V+60)/20}+3e^{-(V+60)/8}} +\frac{50}{3},
\end{equation}
\begin{equation}
    r_{\infty}(V)= \frac{1}{1 + e^{-(v+76)/7}},\\
\end{equation}
\begin{equation}
    \tau_{r}(V)=\frac{10^{5}}{711e^{(V+60)/12}+51e^{-(V+60)/14}} +\frac{25}{3}.
\end{equation}

\begin{acknowledgments}
Mark D. McDonnell's contribution was supported by the Australian Research Council under ARC grant DP1093425 (including an Australian Research Fellowship). 
\end{acknowledgments}

\clearpage

\clearpage

\begin{table}[htdp]
\begin{center}
\begin{tabular}{|c|c|c|c|c|c|c|}
\hline
Current type & $G_{\rm a,det}(t)$ & $\bar{g}_{\rm a}$ (Type II) &  $\bar{g}_{\rm a}$ (Type I-II) & $\bar{g}_{\rm a}$ (Type I-c) & $E_{\rm a}$  \\
\hline
\hline
Sodium (Na) & $m^3h$& 2000 & 2000 & 2000 & +55\\
\hline
High threshold potassium type 1 (KHT1) & $n^2$ & 255 & 255 & 255 & -70 \\
\hline
High threshold potassium type 2 (KHT2) & $p$ & 45 & 45 & 45 & -70 \\
\hline
Low threshold potassium (KLT) & $w^4z$ & 400 & 40 & 0 & -70 \\
\hline
Hyperpolarization-activated cation (h) & $r$ & 40 & 4 & 1 & -43\\
\hline
Leak (lk) & $1$ & 4 & 4 & 4 & -65 \\
\hline
Membrance capacitance, $C_{\rm m}$ & N/A & $12$ & 11.85 & 14.7 & N/A \\
\hline
\end{tabular}
\caption{Summary of parameters for ionic, leak and capacitative currents in the versions of the Rothman-Manis model employed in this paper. The 6 activation variables, $m,h,n,p,w,z$ and $r$  used in the Rothman-Manis model all are time-dependent, and their associated voltage-dependent kinetics equations are given in Appendix A. The maximum conductances given by $\bar{g}_{\rm a}$ have units of nS,  the reversal potentials given by $E_{\rm a}$ have units of mV, the membrane capacitances have units of pF. The reversal potential values are the same as in~\cite{RothmanManis2003b}. The $\bar{g}_{\rm a}$ values for the Type II model are the same as employed by~\cite{Gai10}, which are twice as large as those in~\cite{RothmanManis2003b}, to account for a temperature difference. We therefore also doubled the $\bar{g}_{\rm a}$ values for the Type I-II and Type I-c models, relative to the values stated in~\cite{RothmanManis2003b}.}\label{Table1}
\end{center}
\end{table}

\begin{table}[htdp]
\begin{center}
\begin{tabular}{|c|c|c|c|c|c|c|}
\hline
Ion channel type & Molecular identity& $g_{\rm a}$ (pS) & $N_{\rm a}$ (Type II) &  $N_{\rm a}$ (Type I-II) & $N_{\rm a}$ (Type I-c)   \\
\hline
\hline
Na & unknown &$19-24.9$~\cite{Catterall.05} & 45000 & 45000 & 45000\\
\hline
KHT1 & KCNC1~\cite{RothmanManis}&$27$~\cite{Gutman.05} & 5000 & 5000 & 5000 \\
\hline
KHT2 & KCNC1~\cite{RothmanManis}&$27$~\cite{Gutman.05}  & 1000 & 1000 & 1000 \\
\hline
KLT & KCNA1/2/6~\cite{RothmanManis}&$9-18$~\cite{Gutman.05} & 15000 & 1500 & 0 \\
\hline
h & unknown& unknown~\cite{Hofmann.05};~est.~$20$ & 1000 & 100 & 25 \\
\hline
\end{tabular}
\caption{Estimates of the number of channels of each kind in the Rothman-Manis model. The column labelled $g_{\rm a}$ contains estimates of the single channel conductances taken from the indicated references, based on information about the molecular identity of each channel type in~\cite{RothmanManis}. The estimates of the number of channels of each type, $N_{\rm a}$ in each version of the model were obtained by dividing the mean of $\bar{g}_{\rm a}$ from Table~1 by $N_{\rm a}$, rounding off, and then dividing by two to account for the temperature dependence of total conductance, as described in the main text.}\label{Table2}
\end{center}
\end{table}


\begin{thebibliography}{60}
\expandafter\ifx\csname natexlab\endcsname\relax\def\natexlab#1{#1}\fi
\expandafter\ifx\csname bibnamefont\endcsname\relax
  \def\bibnamefont#1{#1}\fi
\expandafter\ifx\csname bibfnamefont\endcsname\relax
  \def\bibfnamefont#1{#1}\fi
\expandafter\ifx\csname citenamefont\endcsname\relax
  \def\citenamefont#1{#1}\fi
\expandafter\ifx\csname url\endcsname\relax
  \def\url#1{\texttt{#1}}\fi
\expandafter\ifx\csname urlprefix\endcsname\relax\def\urlprefix{URL }\fi
\providecommand{\bibinfo}[2]{#2}
\providecommand{\eprint}[2][]{\url{#2}}

\bibitem[{\citenamefont{Faisal et~al.}(2008)\citenamefont{Faisal, Selen, and
  Wolpert}}]{Faisal.08}
\bibinfo{author}{\bibfnamefont{A.~A.} \bibnamefont{Faisal}},
  \bibinfo{author}{\bibfnamefont{L.~P.~J.} \bibnamefont{Selen}},
  \bibnamefont{and} \bibinfo{author}{\bibfnamefont{D.~M.}
  \bibnamefont{Wolpert}}, \bibinfo{journal}{Nature Reviews Neuroscience}
  \textbf{\bibinfo{volume}{9}}, \bibinfo{pages}{292} (\bibinfo{year}{2008}).

\bibitem[{\citenamefont{Ermentrout et~al.}(2008)\citenamefont{Ermentrout,
  Gal{\'{a}}n, and Urban}}]{Ermentrout.08}
\bibinfo{author}{\bibfnamefont{G.~B.} \bibnamefont{Ermentrout}},
  \bibinfo{author}{\bibfnamefont{R.~F.} \bibnamefont{Gal{\'{a}}n}},
  \bibnamefont{and} \bibinfo{author}{\bibfnamefont{N.~N.} \bibnamefont{Urban}},
  \bibinfo{journal}{Trends in Neurosciences} \textbf{\bibinfo{volume}{31}},
  \bibinfo{pages}{428} (\bibinfo{year}{2008}).

\bibitem[{\citenamefont{Rolls and Deco}(2010)}]{Rolls}
\bibinfo{author}{\bibfnamefont{E.~T.} \bibnamefont{Rolls}} \bibnamefont{and}
  \bibinfo{author}{\bibfnamefont{G.}~\bibnamefont{Deco}},
  \emph{\bibinfo{title}{The Noisy Brain: Stochastic Dynamics as a Principle of
  Brain Function}} (\bibinfo{publisher}{Oxford University Press, USA},
  \bibinfo{year}{2010}).

\bibitem[{\citenamefont{Destexhe and Rudolph-Lilith}(2012)}]{Destexhe}
\bibinfo{author}{\bibfnamefont{A.}~\bibnamefont{Destexhe}} \bibnamefont{and}
  \bibinfo{author}{\bibfnamefont{M.}~\bibnamefont{Rudolph-Lilith}},
  \emph{\bibinfo{title}{Neuronal Noise}} (\bibinfo{publisher}{Springer, New
  York}, \bibinfo{year}{2012}).

\bibitem[{\citenamefont{McDonnell and Ward}(2011)}]{McDonnell}
\bibinfo{author}{\bibfnamefont{M.~D.} \bibnamefont{McDonnell}}
  \bibnamefont{and} \bibinfo{author}{\bibfnamefont{L.~M.} \bibnamefont{Ward}},
  \bibinfo{journal}{Nature Reviews Neuroscience} \textbf{\bibinfo{volume}{12}},
  \bibinfo{pages}{415} (\bibinfo{year}{2011}).

\bibitem[{\citenamefont{Gammaitoni et~al.}(1998)\citenamefont{Gammaitoni,
  H{\"{a}}nggi, Jung, and Marchesoni}}]{Gammaitoni.98}
\bibinfo{author}{\bibfnamefont{L.}~\bibnamefont{Gammaitoni}},
  \bibinfo{author}{\bibfnamefont{P.}~\bibnamefont{H{\"{a}}nggi}},
  \bibinfo{author}{\bibfnamefont{P.}~\bibnamefont{Jung}}, \bibnamefont{and}
  \bibinfo{author}{\bibfnamefont{F.}~\bibnamefont{Marchesoni}},
  \bibinfo{journal}{Reviews of Modern Physics} \textbf{\bibinfo{volume}{70}},
  \bibinfo{pages}{223} (\bibinfo{year}{1998}).

\bibitem[{\citenamefont{McDonnell and Abbott}(2009)}]{McDonnell.PLOS09}
\bibinfo{author}{\bibfnamefont{M.~D.} \bibnamefont{McDonnell}}
  \bibnamefont{and} \bibinfo{author}{\bibfnamefont{D.}~\bibnamefont{Abbott}},
  \bibinfo{journal}{PLoS Computational Biology} \textbf{\bibinfo{volume}{5}},
  \bibinfo{pages}{e1000348} (\bibinfo{year}{2009}).

\bibitem[{\citenamefont{Buesing et~al.}(2011)\citenamefont{Buesing, Bill,
  Nessler, and Maass}}]{Buesing.11}
\bibinfo{author}{\bibfnamefont{L.}~\bibnamefont{Buesing}},
  \bibinfo{author}{\bibfnamefont{J.}~\bibnamefont{Bill}},
  \bibinfo{author}{\bibfnamefont{B.}~\bibnamefont{Nessler}}, \bibnamefont{and}
  \bibinfo{author}{\bibfnamefont{W.}~\bibnamefont{Maass}},
  \bibinfo{journal}{PLoS Computational Biology} \textbf{\bibinfo{volume}{7}},
  \bibinfo{pages}{e1002211} (\bibinfo{year}{2011}).

\bibitem[{\citenamefont{Tsodyks and Markram}(1997)}]{Tsodyks.97}
\bibinfo{author}{\bibfnamefont{M.~V.} \bibnamefont{Tsodyks}} \bibnamefont{and}
  \bibinfo{author}{\bibfnamefont{H.}~\bibnamefont{Markram}},
  \bibinfo{journal}{Proceedings of the National Academy of Sciences of the USA}
  \textbf{\bibinfo{volume}{94}}, \bibinfo{pages}{719} (\bibinfo{year}{1997}).

\bibitem[{\citenamefont{Lindner et~al.}(2009)\citenamefont{Lindner, Gangloff,
  Longtin, and Lewis}}]{Lindner.09}
\bibinfo{author}{\bibfnamefont{B.}~\bibnamefont{Lindner}},
  \bibinfo{author}{\bibfnamefont{D.}~\bibnamefont{Gangloff}},
  \bibinfo{author}{\bibfnamefont{A.}~\bibnamefont{Longtin}}, \bibnamefont{and}
  \bibinfo{author}{\bibfnamefont{J.~E.} \bibnamefont{Lewis}},
  \bibinfo{journal}{The Journal of Neuroscience} \textbf{\bibinfo{volume}{29}},
  \bibinfo{pages}{2076} (\bibinfo{year}{2009}).

\bibitem[{\citenamefont{Hille}(2001)}]{Hille}
\bibinfo{author}{\bibfnamefont{B.}~\bibnamefont{Hille}},
  \emph{\bibinfo{title}{Ion Channels of Excitable Membranes}}
  (\bibinfo{publisher}{Sinauer Associates, Sunderland, MA},
  \bibinfo{year}{2001}), \bibinfo{edition}{3rd} ed.

\bibitem[{\citenamefont{Fox and Lu}(1994)}]{FoxLu}
\bibinfo{author}{\bibfnamefont{R.~F.} \bibnamefont{Fox}} \bibnamefont{and}
  \bibinfo{author}{\bibfnamefont{Y.-N.} \bibnamefont{Lu}},
  \bibinfo{journal}{Physical Review E} \textbf{\bibinfo{volume}{49}},
  \bibinfo{pages}{3421} (\bibinfo{year}{1994}).

\bibitem[{\citenamefont{Fox}(1997)}]{FOX}
\bibinfo{author}{\bibfnamefont{R.~F.} \bibnamefont{Fox}},
  \bibinfo{journal}{Biophysical Journal} \textbf{\bibinfo{volume}{72}},
  \bibinfo{pages}{2068} (\bibinfo{year}{1997}).

\bibitem[{\citenamefont{Chow and White}(1996)}]{Chow}
\bibinfo{author}{\bibfnamefont{C.~C.} \bibnamefont{Chow}} \bibnamefont{and}
  \bibinfo{author}{\bibfnamefont{J.~A.} \bibnamefont{White}},
  \bibinfo{journal}{Biophysical Journal} \textbf{\bibinfo{volume}{71}},
  \bibinfo{pages}{3013} (\bibinfo{year}{1996}).

\bibitem[{\citenamefont{White et~al.}(2000)\citenamefont{White, Rubinstein, and
  Kay}}]{White}
\bibinfo{author}{\bibfnamefont{J.~A.} \bibnamefont{White}},
  \bibinfo{author}{\bibfnamefont{J.~T.} \bibnamefont{Rubinstein}},
  \bibnamefont{and} \bibinfo{author}{\bibfnamefont{A.~R.} \bibnamefont{Kay}},
  \bibinfo{journal}{Trends in Neurosciences} \textbf{\bibinfo{volume}{23}},
  \bibinfo{pages}{131} (\bibinfo{year}{2000}).

\bibitem[{\citenamefont{Schneidman et~al.}(1998)\citenamefont{Schneidman,
  Freedman, and Segev}}]{Schneidman}
\bibinfo{author}{\bibfnamefont{E.}~\bibnamefont{Schneidman}},
  \bibinfo{author}{\bibfnamefont{B.}~\bibnamefont{Freedman}}, \bibnamefont{and}
  \bibinfo{author}{\bibfnamefont{I.}~\bibnamefont{Segev}},
  \bibinfo{journal}{Neural Computation} \textbf{\bibinfo{volume}{10}},
  \bibinfo{pages}{1679} (\bibinfo{year}{1998}).

\bibitem[{\citenamefont{Shuai and Jung}(2005)}]{Shuai.05}
\bibinfo{author}{\bibfnamefont{J.~W.} \bibnamefont{Shuai}} \bibnamefont{and}
  \bibinfo{author}{\bibfnamefont{P.}~\bibnamefont{Jung}},
  \bibinfo{journal}{Physical Review Letters} \textbf{\bibinfo{volume}{95}},
  \bibinfo{pages}{114501} (\bibinfo{year}{2005}).

\bibitem[{\citenamefont{Faisal and Laughlin}(2007)}]{Faisal.07}
\bibinfo{author}{\bibfnamefont{A.~A.} \bibnamefont{Faisal}} \bibnamefont{and}
  \bibinfo{author}{\bibfnamefont{S.~B.} \bibnamefont{Laughlin}},
  \bibinfo{journal}{PLOS Computational Biology} \textbf{\bibinfo{volume}{3}},
  \bibinfo{pages}{e79} (\bibinfo{year}{2007}).

\bibitem[{\citenamefont{Goldwyn et~al.}(2007)\citenamefont{Goldwyn, Rubinstein,
  and Shea-Brown}}]{Rowat.07}
\bibinfo{author}{\bibfnamefont{J.~H.} \bibnamefont{Goldwyn}},
  \bibinfo{author}{\bibfnamefont{J.~T.} \bibnamefont{Rubinstein}},
  \bibnamefont{and}
  \bibinfo{author}{\bibfnamefont{E.}~\bibnamefont{Shea-Brown}},
  \bibinfo{journal}{Neural Computation} \textbf{\bibinfo{volume}{19}},
  \bibinfo{pages}{1215} (\bibinfo{year}{2007}).

\bibitem[{\citenamefont{Ashida and Kubo}(2010)}]{Ashida}
\bibinfo{author}{\bibfnamefont{G.}~\bibnamefont{Ashida}} \bibnamefont{and}
  \bibinfo{author}{\bibfnamefont{M.}~\bibnamefont{Kubo}},
  \bibinfo{journal}{Physica D} \textbf{\bibinfo{volume}{239}},
  \bibinfo{pages}{327} (\bibinfo{year}{2010}).

\bibitem[{\citenamefont{Goldwyn and Shea-Brown}(2011)}]{Goldwynwhat}
\bibinfo{author}{\bibfnamefont{J.~H.} \bibnamefont{Goldwyn}} \bibnamefont{and}
  \bibinfo{author}{\bibfnamefont{E.}~\bibnamefont{Shea-Brown}},
  \bibinfo{journal}{PLoS Computational Biology} \textbf{\bibinfo{volume}{7}},
  \bibinfo{pages}{041908} (\bibinfo{year}{2011}).

\bibitem[{\citenamefont{Bezrukov and Voydanoy}(1995)}]{Bezrukov.95}
\bibinfo{author}{\bibfnamefont{S.~M.} \bibnamefont{Bezrukov}} \bibnamefont{and}
  \bibinfo{author}{\bibfnamefont{I.}~\bibnamefont{Voydanoy}},
  \bibinfo{journal}{Nature} \textbf{\bibinfo{volume}{378}},
  \bibinfo{pages}{362} (\bibinfo{year}{1995}).

\bibitem[{\citenamefont{Goychuk and H{\"{a}}nggi}(2000)}]{Goychuk.00}
\bibinfo{author}{\bibfnamefont{I.}~\bibnamefont{Goychuk}} \bibnamefont{and}
  \bibinfo{author}{\bibfnamefont{P.}~\bibnamefont{H{\"{a}}nggi}},
  \bibinfo{journal}{Physical Review E} \textbf{\bibinfo{volume}{61}},
  \bibinfo{pages}{2272} (\bibinfo{year}{2000}).

\bibitem[{\citenamefont{Bruce}(2007)}]{Bruce.07}
\bibinfo{author}{\bibfnamefont{I.~C.} \bibnamefont{Bruce}},
  \bibinfo{journal}{Annals of Biomedical Engineering}
  \textbf{\bibinfo{volume}{35}}, \bibinfo{pages}{315} (\bibinfo{year}{2007}).

\bibitem[{\citenamefont{Bruce}(2009)}]{Bruce.09}
\bibinfo{author}{\bibfnamefont{I.~C.} \bibnamefont{Bruce}},
  \bibinfo{journal}{Annals of Biomedical Engineering}
  \textbf{\bibinfo{volume}{37}}, \bibinfo{pages}{824} (\bibinfo{year}{2009}).

\bibitem[{\citenamefont{Sengupta et~al.}(2010)\citenamefont{Sengupta, Laughlin,
  and Niven}}]{Sengupta.10}
\bibinfo{author}{\bibfnamefont{B.}~\bibnamefont{Sengupta}},
  \bibinfo{author}{\bibfnamefont{S.~B.} \bibnamefont{Laughlin}},
  \bibnamefont{and} \bibinfo{author}{\bibfnamefont{J.~E.} \bibnamefont{Niven}},
  \bibinfo{journal}{Physical Review E} \textbf{\bibinfo{volume}{81}},
  \bibinfo{pages}{011918} (\bibinfo{year}{2010}).

\bibitem[{\citenamefont{Goldwyn et~al.}(2011)\citenamefont{Goldwyn, Imennov,
  Famulare, and Shea-Brown}}]{Goldwyn}
\bibinfo{author}{\bibfnamefont{J.~H.} \bibnamefont{Goldwyn}},
  \bibinfo{author}{\bibfnamefont{N.~S.} \bibnamefont{Imennov}},
  \bibinfo{author}{\bibfnamefont{M.}~\bibnamefont{Famulare}}, \bibnamefont{and}
  \bibinfo{author}{\bibfnamefont{E.}~\bibnamefont{Shea-Brown}},
  \bibinfo{journal}{Physical Review E} \textbf{\bibinfo{volume}{83}},
  \bibinfo{pages}{041908} (\bibinfo{year}{2011}).

\bibitem[{\citenamefont{Linaro et~al.}(2011)\citenamefont{Linaro, Storace, and
  Giugliano}}]{Linaro}
\bibinfo{author}{\bibfnamefont{D.}~\bibnamefont{Linaro}},
  \bibinfo{author}{\bibfnamefont{M.}~\bibnamefont{Storace}}, \bibnamefont{and}
  \bibinfo{author}{\bibfnamefont{M.}~\bibnamefont{Giugliano}},
  \bibinfo{journal}{PLoS Computational Biology} \textbf{\bibinfo{volume}{7}},
  \bibinfo{pages}{e1001102} (\bibinfo{year}{2011}).

\bibitem[{\citenamefont{Schmandt and Gal{\'{a}}n}(2012)}]{Schmandt}
\bibinfo{author}{\bibfnamefont{N.~T.} \bibnamefont{Schmandt}} \bibnamefont{and}
  \bibinfo{author}{\bibfnamefont{R.~F.} \bibnamefont{Gal{\'{a}}n}},
  \bibinfo{journal}{Physical Review Letters} \textbf{\bibinfo{volume}{109}},
  \bibinfo{pages}{118101} (\bibinfo{year}{2012}).

\bibitem[{\citenamefont{Dangerfield et~al.}(2012)\citenamefont{Dangerfield,
  Kay, and Burrage}}]{Dangerfield}
\bibinfo{author}{\bibfnamefont{C.~E.} \bibnamefont{Dangerfield}},
  \bibinfo{author}{\bibfnamefont{D.}~\bibnamefont{Kay}}, \bibnamefont{and}
  \bibinfo{author}{\bibfnamefont{K.}~\bibnamefont{Burrage}},
  \bibinfo{journal}{Physical Review E} \textbf{\bibinfo{volume}{85}},
  \bibinfo{pages}{051907} (\bibinfo{year}{2012}).

\bibitem[{\citenamefont{Orio and Soudry}(2012)}]{Orio.12}
\bibinfo{author}{\bibfnamefont{P.}~\bibnamefont{Orio}} \bibnamefont{and}
  \bibinfo{author}{\bibfnamefont{D.}~\bibnamefont{Soudry}},
  \bibinfo{journal}{PLoS One} \textbf{\bibinfo{volume}{7}},
  \bibinfo{pages}{e36670} (\bibinfo{year}{2012}).

\bibitem[{\citenamefont{Huang et~al.}(2013)\citenamefont{Huang, R{\"{u}}diger,
  and Shuai}}]{Huang.13}
\bibinfo{author}{\bibfnamefont{Y.}~\bibnamefont{Huang}},
  \bibinfo{author}{\bibfnamefont{S.}~\bibnamefont{R{\"{u}}diger}},
  \bibnamefont{and} \bibinfo{author}{\bibfnamefont{J.}~\bibnamefont{Shuai}},
  \bibinfo{journal}{Physical Review E} \textbf{\bibinfo{volume}{87}},
  \bibinfo{pages}{012716} (\bibinfo{year}{2013}).

\bibitem[{\citenamefont{Mino et~al.}(2002)\citenamefont{Mino, Rubinstein, and
  White}}]{Mino02}
\bibinfo{author}{\bibfnamefont{H.}~\bibnamefont{Mino}},
  \bibinfo{author}{\bibfnamefont{J.~T.} \bibnamefont{Rubinstein}},
  \bibnamefont{and} \bibinfo{author}{\bibfnamefont{J.~A.} \bibnamefont{White}},
  \bibinfo{journal}{Annals of Biomedical Engineering}
  \textbf{\bibinfo{volume}{30}}, \bibinfo{pages}{578} (\bibinfo{year}{2002}).

\bibitem[{\citenamefont{Mino et~al.}(2004)\citenamefont{Mino, Rubinstein,
  Miller, and Abbas}}]{Mino04}
\bibinfo{author}{\bibfnamefont{H.}~\bibnamefont{Mino}},
  \bibinfo{author}{\bibfnamefont{J.~T.} \bibnamefont{Rubinstein}},
  \bibinfo{author}{\bibfnamefont{C.~A.} \bibnamefont{Miller}},
  \bibnamefont{and} \bibinfo{author}{\bibfnamefont{P.~J.} \bibnamefont{Abbas}},
  \bibinfo{journal}{IEEE Transactions on Biomedical Engineering}
  \textbf{\bibinfo{volume}{51}}, \bibinfo{pages}{13} (\bibinfo{year}{2004}).

\bibitem[{\citenamefont{Imennov and Rubinstein}(2009)}]{Imennov}
\bibinfo{author}{\bibfnamefont{N.~S.} \bibnamefont{Imennov}} \bibnamefont{and}
  \bibinfo{author}{\bibfnamefont{J.~T.} \bibnamefont{Rubinstein}},
  \bibinfo{journal}{IEEE Transactions on Biomedical Engineering}
  \textbf{\bibinfo{volume}{56}}, \bibinfo{pages}{2493} (\bibinfo{year}{2009}).

\bibitem[{\citenamefont{Goldwyn et~al.}(2012)\citenamefont{Goldwyn, Rubinstein,
  and Shea-Brown}}]{Goldwyn.12}
\bibinfo{author}{\bibfnamefont{J.~H.} \bibnamefont{Goldwyn}},
  \bibinfo{author}{\bibfnamefont{J.~T.} \bibnamefont{Rubinstein}},
  \bibnamefont{and}
  \bibinfo{author}{\bibfnamefont{E.}~\bibnamefont{Shea-Brown}},
  \bibinfo{journal}{Journal of Neurophysiology} \textbf{\bibinfo{volume}{108}},
  \bibinfo{pages}{1430} (\bibinfo{year}{2012}).

\bibitem[{\citenamefont{Rothman and Manis.}(2003{\natexlab{a}})}]{RothmanManis}
\bibinfo{author}{\bibfnamefont{J.~S.} \bibnamefont{Rothman}} \bibnamefont{and}
  \bibinfo{author}{\bibfnamefont{P.~B.} \bibnamefont{Manis.}},
  \bibinfo{journal}{Journal of Neurophysiology} \textbf{\bibinfo{volume}{89}},
  \bibinfo{pages}{3083} (\bibinfo{year}{2003}{\natexlab{a}}).

\bibitem[{\citenamefont{Rothman and
  Manis.}(2003{\natexlab{b}})}]{RothmanManis2003a}
\bibinfo{author}{\bibfnamefont{J.~S.} \bibnamefont{Rothman}} \bibnamefont{and}
  \bibinfo{author}{\bibfnamefont{P.~B.} \bibnamefont{Manis.}},
  \bibinfo{journal}{Journal of Neurophysiology} \textbf{\bibinfo{volume}{89}},
  \bibinfo{pages}{3070} (\bibinfo{year}{2003}{\natexlab{b}}).

\bibitem[{\citenamefont{Rothman and
  Manis.}(2003{\natexlab{c}})}]{RothmanManis2003b}
\bibinfo{author}{\bibfnamefont{J.~S.} \bibnamefont{Rothman}} \bibnamefont{and}
  \bibinfo{author}{\bibfnamefont{P.~B.} \bibnamefont{Manis.}},
  \bibinfo{journal}{Journal of Neurophysiology} \textbf{\bibinfo{volume}{89}},
  \bibinfo{pages}{3097} (\bibinfo{year}{2003}{\natexlab{c}}).

\bibitem[{\citenamefont{Gai et~al.}(2009)\citenamefont{Gai, Doiron, Kotak, and
  Rinzel}}]{Gai09}
\bibinfo{author}{\bibfnamefont{Y.}~\bibnamefont{Gai}},
  \bibinfo{author}{\bibfnamefont{B.}~\bibnamefont{Doiron}},
  \bibinfo{author}{\bibfnamefont{V.}~\bibnamefont{Kotak}}, \bibnamefont{and}
  \bibinfo{author}{\bibfnamefont{J.}~\bibnamefont{Rinzel}},
  \bibinfo{journal}{Journal of Neurophysiology} \textbf{\bibinfo{volume}{102}},
  \bibinfo{pages}{3447} (\bibinfo{year}{2009}).

\bibitem[{\citenamefont{Gai et~al.}(2010)\citenamefont{Gai, Doiron, and
  Rinzel}}]{Gai10}
\bibinfo{author}{\bibfnamefont{Y.}~\bibnamefont{Gai}},
  \bibinfo{author}{\bibfnamefont{B.}~\bibnamefont{Doiron}}, \bibnamefont{and}
  \bibinfo{author}{\bibfnamefont{J.}~\bibnamefont{Rinzel}},
  \bibinfo{journal}{PLOS Computational Biology} \textbf{\bibinfo{volume}{6}},
  \bibinfo{pages}{e10000825} (\bibinfo{year}{2010}).

\bibitem[{\citenamefont{Gutkin et~al.}(2009)\citenamefont{Gutkin, Jost, and
  Tuckwell}}]{Gutkin}
\bibinfo{author}{\bibfnamefont{B.~S.} \bibnamefont{Gutkin}},
  \bibinfo{author}{\bibfnamefont{J.}~\bibnamefont{Jost}}, \bibnamefont{and}
  \bibinfo{author}{\bibfnamefont{H.~C.} \bibnamefont{Tuckwell}},
  \bibinfo{journal}{Naturwissenschaften} \textbf{\bibinfo{volume}{96}},
  \bibinfo{pages}{1091} (\bibinfo{year}{2009}).

\bibitem[{\citenamefont{Tuckwell et~al.}(2009)\citenamefont{Tuckwell, Jost, and
  Gutkin}}]{Tuckwell.09}
\bibinfo{author}{\bibfnamefont{H.~C.} \bibnamefont{Tuckwell}},
  \bibinfo{author}{\bibfnamefont{J.}~\bibnamefont{Jost}}, \bibnamefont{and}
  \bibinfo{author}{\bibfnamefont{B.~S.} \bibnamefont{Gutkin}},
  \bibinfo{journal}{Physical Review E} \textbf{\bibinfo{volume}{80}},
  \bibinfo{pages}{031907} (\bibinfo{year}{2009}).

\bibitem[{\citenamefont{Tuckwell and Jost}(2010)}]{Tuckwell.10}
\bibinfo{author}{\bibfnamefont{H.~C.} \bibnamefont{Tuckwell}} \bibnamefont{and}
  \bibinfo{author}{\bibfnamefont{J.}~\bibnamefont{Jost}},
  \bibinfo{journal}{PLoS Computational Biology} \textbf{\bibinfo{volume}{6}},
  \bibinfo{pages}{e1000794} (\bibinfo{year}{2010}).

\bibitem[{\citenamefont{Tuckwell and Jost}(2012)}]{Tuckwell}
\bibinfo{author}{\bibfnamefont{H.~C.} \bibnamefont{Tuckwell}} \bibnamefont{and}
  \bibinfo{author}{\bibfnamefont{J.}~\bibnamefont{Jost}},
  \bibinfo{journal}{Physica A} \textbf{\bibinfo{volume}{391}},
  \bibinfo{pages}{5311} (\bibinfo{year}{2012}).

\bibitem[{\citenamefont{Paydarfar et~al.}(2006)\citenamefont{Paydarfar, Forger,
  and Clay}}]{Paydarfar.06}
\bibinfo{author}{\bibfnamefont{D.}~\bibnamefont{Paydarfar}},
  \bibinfo{author}{\bibfnamefont{D.~B.} \bibnamefont{Forger}},
  \bibnamefont{and} \bibinfo{author}{\bibfnamefont{J.~R.} \bibnamefont{Clay}},
  \bibinfo{journal}{Journal of Neurophysiology} \textbf{\bibinfo{volume}{96}},
  \bibinfo{pages}{3338} (\bibinfo{year}{2006}).

\bibitem[{\citenamefont{Hodgkin and Huxley}(1952)}]{HH}
\bibinfo{author}{\bibfnamefont{A.~L.} \bibnamefont{Hodgkin}} \bibnamefont{and}
  \bibinfo{author}{\bibfnamefont{A.~F.} \bibnamefont{Huxley}},
  \bibinfo{journal}{Journal of Physiology} \textbf{\bibinfo{volume}{117}},
  \bibinfo{pages}{500} (\bibinfo{year}{1952}).

\bibitem[{\citenamefont{Izhikevich}(2004)}]{Izhikevich.04}
\bibinfo{author}{\bibfnamefont{E.~M.} \bibnamefont{Izhikevich}},
  \bibinfo{journal}{IEEE Transactions on Neural Networks}
  \textbf{\bibinfo{volume}{15}}, \bibinfo{pages}{1063} (\bibinfo{year}{2004}).

\bibitem[{\citenamefont{Rothman et~al.}(1993)\citenamefont{Rothman, Young, and
  Manis.}}]{Rothman93}
\bibinfo{author}{\bibfnamefont{J.~S.} \bibnamefont{Rothman}},
  \bibinfo{author}{\bibfnamefont{E.~D.} \bibnamefont{Young}}, \bibnamefont{and}
  \bibinfo{author}{\bibfnamefont{P.~B.} \bibnamefont{Manis.}},
  \bibinfo{journal}{J Neurophysiol} \textbf{\bibinfo{volume}{70}},
  \bibinfo{pages}{2562} (\bibinfo{year}{1993}).

\bibitem[{\citenamefont{Manis and Marx.}(1991)}]{ManisMarx}
\bibinfo{author}{\bibfnamefont{P.~B.} \bibnamefont{Manis}} \bibnamefont{and}
  \bibinfo{author}{\bibfnamefont{S.~O.} \bibnamefont{Marx.}},
  \bibinfo{journal}{The Journal of Neuroscience} \textbf{\bibinfo{volume}{11}},
  \bibinfo{pages}{2865} (\bibinfo{year}{1991}).

\bibitem[{\citenamefont{Dayan and Abbott}(2005)}]{dayan}
\bibinfo{author}{\bibfnamefont{P.}~\bibnamefont{Dayan}} \bibnamefont{and}
  \bibinfo{author}{\bibfnamefont{L.~F.} \bibnamefont{Abbott}},
  \emph{\bibinfo{title}{Theoretical Neuroscience Computational and Mathematical
  Modeling of Neural Systems}} (\bibinfo{publisher}{The MIT Press},
  \bibinfo{year}{2005}).

\bibitem[{\citenamefont{Pakdaman et~al.}(2010)\citenamefont{Pakdaman,
  Thieullen, and Wainrib}}]{Pakdaman}
\bibinfo{author}{\bibfnamefont{K.}~\bibnamefont{Pakdaman}},
  \bibinfo{author}{\bibfnamefont{M.}~\bibnamefont{Thieullen}},
  \bibnamefont{and} \bibinfo{author}{\bibfnamefont{G.}~\bibnamefont{Wainrib}},
  \bibinfo{journal}{Adv Appl Probab} \textbf{\bibinfo{volume}{42}},
  \bibinfo{pages}{761} (\bibinfo{year}{2010}).

\bibitem[{\citenamefont{Kloeden and Platen}(1992)}]{Kloeden}
\bibinfo{author}{\bibfnamefont{P.~E.} \bibnamefont{Kloeden}} \bibnamefont{and}
  \bibinfo{author}{\bibfnamefont{E.}~\bibnamefont{Platen}},
  \emph{\bibinfo{title}{Numerical Solution of Stochastic Differential
  Equations}} (\bibinfo{publisher}{Springer-Verlag}, \bibinfo{year}{1992}).

\bibitem[{\citenamefont{Stocks}(2000)}]{Stocks.Mar2000}
\bibinfo{author}{\bibfnamefont{N.~G.} \bibnamefont{Stocks}},
  \bibinfo{journal}{Physical Review Letters} \textbf{\bibinfo{volume}{84}},
  \bibinfo{pages}{2310} (\bibinfo{year}{2000}).

\bibitem[{\citenamefont{McDonnell et~al.}(2007)\citenamefont{McDonnell, Stocks,
  and Abbott}}]{McDonnell.07}
\bibinfo{author}{\bibfnamefont{M.~D.} \bibnamefont{McDonnell}},
  \bibinfo{author}{\bibfnamefont{N.~G.} \bibnamefont{Stocks}},
  \bibnamefont{and} \bibinfo{author}{\bibfnamefont{D.}~\bibnamefont{Abbott}},
  \bibinfo{journal}{Physical Review E} \textbf{\bibinfo{volume}{75}},
  \bibinfo{pages}{061105} (\bibinfo{year}{2007}).

\bibitem[{\citenamefont{Izhikevich}(2007)}]{Izhikevich}
\bibinfo{author}{\bibfnamefont{E.~M.} \bibnamefont{Izhikevich}},
  \emph{\bibinfo{title}{Dynamical Systems in Neuroscience}}
  (\bibinfo{publisher}{The MIT Press, Cambridge, MA}, \bibinfo{year}{2007}).

\bibitem[{\citenamefont{Buchholtz et~al.}(2002)\citenamefont{Buchholtz,
  Schinor, and Schneider}}]{Buchholtz.02}
\bibinfo{author}{\bibfnamefont{F.}~\bibnamefont{Buchholtz}},
  \bibinfo{author}{\bibfnamefont{N.}~\bibnamefont{Schinor}}, \bibnamefont{and}
  \bibinfo{author}{\bibfnamefont{F.~W.} \bibnamefont{Schneider}},
  \bibinfo{journal}{Journal of Physical Chemistry B}
  \textbf{\bibinfo{volume}{106}}, \bibinfo{pages}{5086} (\bibinfo{year}{2002}).

\bibitem[{\citenamefont{Gutman and {\em et. al}}(2005)}]{Gutman.05}
\bibinfo{author}{\bibfnamefont{G.~A.} \bibnamefont{Gutman}} \bibnamefont{and}
  \bibinfo{author}{\bibnamefont{{\em et. al}}},
  \bibinfo{journal}{Pharmacological Reviews} \textbf{\bibinfo{volume}{57}},
  \bibinfo{pages}{473} (\bibinfo{year}{2005}).

\bibitem[{\citenamefont{Hofmann et~al.}(2005)\citenamefont{Hofmann, Biel, and
  Kaupp}}]{Hofmann.05}
\bibinfo{author}{\bibfnamefont{F.}~\bibnamefont{Hofmann}},
  \bibinfo{author}{\bibfnamefont{M.}~\bibnamefont{Biel}}, \bibnamefont{and}
  \bibinfo{author}{\bibfnamefont{U.~B.} \bibnamefont{Kaupp}},
  \bibinfo{journal}{Pharmacological Reviews} \textbf{\bibinfo{volume}{57}},
  \bibinfo{pages}{455} (\bibinfo{year}{2005}).

\bibitem[{\citenamefont{Catterall et~al.}(2005)\citenamefont{Catterall, Goldin,
  and Waxman}}]{Catterall.05}
\bibinfo{author}{\bibfnamefont{W.~A.} \bibnamefont{Catterall}},
  \bibinfo{author}{\bibfnamefont{A.~L.} \bibnamefont{Goldin}},
  \bibnamefont{and} \bibinfo{author}{\bibfnamefont{S.~G.}
  \bibnamefont{Waxman}}, \bibinfo{journal}{Pharmacological Reviews}
  \textbf{\bibinfo{volume}{57}}, \bibinfo{pages}{397} (\bibinfo{year}{2005}).

\end{thebibliography}
\end{document}